\documentclass[twocolumn]{aastex631}
\usepackage{hyperref}
\usepackage[caption=false]{subfig}
\usepackage{booktabs}
\usepackage{censor}
\usepackage[arrowdel]{physics}
\usepackage{mathtools}
\usepackage{outlines}
\usepackage{lipsum}  
\usepackage{amsmath,bm}

\definecolor{belize}{RGB}{41, 128, 185}
\definecolor{peter}{RGB}{52, 152, 219}
\definecolor{nephritis}{RGB}{39, 174, 96}
\definecolor{asbestos}{RGB}{127, 140, 141}
\definecolor{clouds}{RGB}{236, 240, 241}

\hypersetup{linkcolor=belize, citecolor=belize,filecolor=asbestos,urlcolor=peter}

\DeclareMathOperator{\resample}{resample}

\def\kmps{\mathrm{km}\;\mathrm{s}^{-1}}

\begin{document}
\shorttitle{\emph{blas\'e}: Interpretable Machine Learning for Spectroscopy}
\shortauthors{Gully-Santiago \& Morley}
\title{An Interpretable Machine Learning Framework for Modeling High-Resolution Spectroscopic Data\footnote{Open source code at \url{https://github.com/gully/blase}}}

\author{Michael Gully-Santiago}
\affiliation{The University of Texas at Austin Department of Astronomy}

\author{Caroline V. Morley}
\affiliation{The University of Texas at Austin Department of Astronomy}

\begin{abstract}

    Comparison of \'echelle spectra to synthetic models has become a computational statistics challenge, with over ten thousand individual spectral lines affecting a typical cool star \'echelle spectrum.  Telluric artifacts, imperfect line lists, inexact continuum placement, and inflexible models frustrate the scientific promise of these information-rich datasets.  Here we debut an interpretable machine-learning framework \emph{blas\'e} that addresses these and other challenges.  The semi-empirical approach can be viewed as ``transfer learning''--first pre-training models on noise-free precomputed synthetic spectral models, then learning the corrections to line depths and widths from whole-spectrum fitting to an observed spectrum.  The auto-differentiable model employs back-propagation, the fundamental algorithm empowering modern Deep Learning and Neural Networks. Here, however, the 40,000+ parameters symbolize physically interpretable line profile properties such as amplitude, width, location, and shape, plus radial velocity and rotational broadening.  This hybrid data-/model- driven framework allows joint modeling of stellar and telluric lines simultaneously, a potentially transformative step forwards for mitigating the deleterious telluric contamination in the near-infrared.  The \emph{blas\'e} approach acts as both a deconvolution tool and semi-empirical model. The general purpose scaffolding may be extensible to many scientific applications, including Precision Radial Velocities, Doppler Imaging, chemical abundances for Galactic archaeology, line veiling, magnetic fields, and remote sensing.  Its sparse-matrix architecture and GPU-acceleration make \emph{blas\'e} fast.  The open-source PyTorch-based code \texttt{blase} includes tutorials, Application Programming Interface (API) documentation, and more.  We show how the tool fits into the existing Python spectroscopy ecosystem, demonstrate a range of astrophysical applications, and discuss limitations and future extensions.

\end{abstract}


\section{Introduction}\label{sec:intro}

\subsection{Spectral fitting past and present}

Tens of thousands or more individual spectral lines give rise to a sea of undulations that imbue each stellar spectrum with its characteristic appearance.  The identification and understanding of these lines have defined a large category of astrophysics over the last century.  The field grew from by-eye catalogs of stellar templates \citep{1901AnHar..28..129C} to quantifying the role of atomic ionization balance \citep{1925PhDT.........1P}, to modern synthetic forward models including millions or billions of lines \citep[\emph{e.g.}][]{husser13, 2021ApJ...920...85M, 2021JQSRT.26107476V}.  As technology has improved, our data and models have become more voluminous, more precise, and more complicated.  The mere act of comparing models to observed spectra can now resemble a computational statistics challenge as much as a scientific one.  Here we introduce a new machine-learning-based framework \emph{blas\'e} aimed at solving computational, statistical, and scientific challenges associated with data-model comparisons for modern astronomical spectroscopy.

The metaphorical holy grail of astronomical spectroscopy would be a function that takes in an observed stellar spectrum and reports back the position, amplitude, width, and shape of all of its spectral lines, automatically, accurately, and precisely.  The function would go further. It would report back the systemic radial velocity ($RV$) and rotational broadening ($v\sin{i}$) and fundamental stellar properties, including $T_{\mathrm{eff}}$, $\log{g}$, and $[\mathrm{Fe}/\mathrm{H}]$.  Finally, the function would---in a feat of artificial intelligence---provide what it believes to be the interpretable generating function that produced this data in the first place, so that we may gain insights on future examples of this or other stars.  Solving this problem is hard, for at least four reasons.  First, the spectral lines may overlap, and so the assignment of one line may be partially degenerate with the assignment of some other adjacent line.  Second, extremely wide line wings blend into the continuum, such that the placement of the continuum level may become ill-defined.  Third, the extent of line blending and realized line shape depends strongly on the spectral resolution of the spectrograph, the rotational broadening of the star, and possibly the instrumental configuration at the time of observation.  Finally, telluric absorption lines commingle with the astronomical spectral lines of interest, censoring some spectral regions entirely, or partially confounding other lines with chance alignments.

Addressing these and other challenges forms the backbone of \emph{spectral calibration}, an increasingly valuable specialty as the deficits in our models become intolerable with greater data quantity and quality.  Luckily, many scientific applications in astrophysics do not need the technically demanding noise-free template, nor catalog-of-all-spectral-lines.  A few lines suffice.  For those applications, human inspection of isolated lines and semi-automated equivalent width determination have been---and will remain---adequate.

But many new and important questions in the fields of stars and exoplanets aspire to reach the margins of what the entire dataset can inform.  In particular, data from high-\emph{grasp} \'echelle spectrographs possess simultaneously high spectral resolving power and high bandwidth, yielding tens of thousands or possibly millions of independent spectral resolution elements for each star, substar, or exoplanet.  Scientific applications that seek to gain signal by ``stacking'' spectral lines or cross-correlating with templates can hypothetically gain huge boosts in the accessible signal-to-noise ratio compared to a single or few lines.  Most manual and semi-automated methods cannot take advantage of the entire spectral bandwidth, or rely on exact knowledge of the underlying templates and may fail to achieve the hypothetical promise of these high-bandwidth spectrographs \citep{2020AJ....160..198H}.

For example, exoplanet cross-correlation spectroscopy \citep{2010Natur.465.1049S,2012Natur.486..502B,2013MNRAS.436L..35B} hinges on accurate molecular spectral templates to detect and characterize the atmospheres of exoplanets.  Imperfections in these templates can mute the perceived signal strength of these atmospheric features \citep{2015A&A...575A..20H}.

In extreme precision radial velocity (EPRV) applications, cross-correlation methods work \citep{2018A_A...620A..47D}, but have many limitations \citep{2022arXiv220110639Z}.  Among the many such limitations, one pernicious noise floor stands out as enigmatic: telluric mitigation.  Many practitioners today simply mask these telluric regions, yet micro-tellurics still inject variance into the spectrum that cannot be easily accounted for with existing methods.  Instead, a robust accounting of telluric absorption at the $\mathrm{cm/s}$ level may require joint modeling of the star and the Earth's atmospheric absorption \emph{before} convolution with an instrumental kernel.  This telluric joint modeling capability does not yet exist at a precision that can meet these strenuous demands.

In the case of Doppler imaging, an accurate underlying spectral template is needed to detect longitudinally symmetric structures \citep{1983PASP...95..565V,2021arXiv211006271L} such as polar spots \citep{roettenbacher16} or zonal bands \citep{Crossfield14,2021ApJ...906...64A}.  There exists a nearly circular reasoning: we need to know the extent of line profile perturbations to reveal the underlying spectral template, but we need the underlying spectral template to estimate the line profile perturbations.  The approaches introduced here offer a path forward on these long-standing friction points.

\subsection{Automatic differentiation technology}

Existing open-source frameworks have overcome some of these challenges, or have been purpose-built for specialized applications.
These frameworks include \texttt{ROBOSPECT} \citep{2013PASP..125.1164W}, \texttt{specmatch} \citep{2015PhDT........82P}, \texttt{specmatch-emp} \citep{2017ApJ...836...77Y}, \texttt{wobble} \citep{2019AJ....158..164B}, \texttt{starfish} \citep{czekala15}, \texttt{sick} \citep{2016ApJS..223....8C}, \texttt{psoap} \citep{2017ApJ...840...49C}, \texttt{FAL} (Cargile et al. \emph{in prep}), CHIMERA \citep{2015ApJ...807..183L}, the \texttt{Cannon} \citep{2017ApJ...836....5H},  \texttt{MOOG} \citep{2012ascl.soft02009S}, \texttt{MOOGStokes} \citep{2013AJ....146...51D}, \texttt{MINESweeper} \citep{2020ApJ...900...28C}, and recently \texttt{ExoJAX} \citep{2022ApJS..258...31K}.
The designs of these frameworks necessarily have to make a choice in the bias-variance tradeoff: is the tool more \emph{data}-driven or more \emph{model}-driven?  The statistical tradeoff can be viewed as a concession in physical self-consistency for model flexibility: more or fewer parameters; more \emph{accurate} or more \emph{precise}.

A key new enabling technology breaks these classical tradeoffs in data-model comparisons for astronomical spectroscopy.  Automatic differentiation \citep[``autodiff'' or ``autograd'',][]{2015arXiv150205767G, 2016PhDT.......317M} and its affiliated backpropagation algorithm \citep{kelley1960,Linnainmaa1976TaylorEO, 1986Natur.323..533R, 1990JGCD...13..926D} has revolutionized machine learning and neural network architecture design, and is increasingly applied in astrophysical data analysis contexts, \emph{e.g.} kernel phase coronography with \texttt{morphine} \citep{2021ApJ...907...40P,2021JOSAB..38.2465W} and $\partial$\texttt{lux} \citep{10.1117/12.2629774}, stellar surface modeling with \texttt{starry} \citep{2021AJ....162..123L}, and exoplanet orbit fitting with \texttt{exoplanet} \citep{2021JOSS....6.3285F}.  Of the spectroscopy frameworks mentioned above, the \texttt{TensorFlow}-based \citep{tensorflow2015-whitepaper} \texttt{wobble} and the \texttt{JAX}-based \citep{jax2018github} \texttt{ExoJAX} employ autodiff technology.  \texttt{wobble} treats each pixel as a tunable control point, producing $\sim10^5$ parameters for a modern stellar spectrum.  The \texttt{ExoJAX} framework has only $\sim$100 tunable parameters that describe the fundamental physical properties controlling a stratified brown dwarf atmosphere.  These two autodiff-aware frameworks span the extreme ends of non-parametric and parametric modeling for spectroscopy.

In this paper, we show that autodiff-aware semi-empirical models offer an appealing middle ground: informed by self-consistent models but refined with data.  This sweet spot in the bias-variance tradeoff can be thought of as a hybrid data-and-model driven approach.  The algorithm presented here focuses on modeling the spectra of stars and brown dwarfs.  Existing stellar models \citep[\emph{e.g.}][]{husser13} and substellar models \citep[\emph{e.g.}][]{2021ApJ...920...85M} laboriously solve for a self-consistent thermal structure in the atmosphere given the copious opacity sources that themselves depend on temperature and pressure.  Here we build upon that hard work by cloning pre-existing synthetic stellar or substellar models (Section \ref{methodology}), and optionally by cloning models of Earth's atmospheric ``telluric'' absorption (Section \ref{sectionTelluric}). We introduce the interpretable forward-model design and its \texttt{PyTorch}-based \citep{2019arXiv191201703P} implementation, \texttt{blase}.  In Section \ref{transferLearn}  we describe how to adapt both stellar and telluric cloned models simultaneously, using a transfer-learning step.  We obtain semi-empirical models by comparing to real-world \'echelle data in Section \ref{secResults}. Finally, we discuss perspectives on how to think of \emph{blas\'e} (Section \ref{secDiscuss}) and chronicle many conceivable extensions for unlocking new science (Section \ref{secFutureWork}).

\section{Methodology I: Cloning stellar spectra}\label{methodology}

Our first stage is to \emph{clone} a precomputed synthetic spectrum.  Cloning means to mimic, emulate, or otherwise approximate the appearance of a discretely sampled spectrum with some function or combination of functions.  The subject of the cloning shall be a noise-free spectrum generated from physics-based atmosphere models.  The purpose of cloning may appear superficial at first glance, but it serves as a necessary gateway for our ultimate data-model comparison goals.

\subsection{Overall Architecture and Design Choices}

We start with a high resolution pre-computed synthetic stellar or substellar model spectrum, $\mathsf{S}_{\rm abs}(\bm{\lambda})$ at its native resolution sampling and with its original absolute physical flux units. The procedure that follows is largely agnostic to the exact details of how this spectrum was made, or what physics or chemistry it may represent. For the purposes of this paper, we will showcase examples from two well-known families of precomputed synthetic astronomical spectra: \texttt{PHOENIX} \citep{husser13} for stellar spectra $(T_{\mathrm{eff}}\in [2300, 10000]\;K)$ and \texttt{Sonora} \citep{2021ApJ...920...85M} for substellar spectra $(T_{\mathrm{eff}}\in [200, 2300]\;K)$. The algorithms in the framework may also work for precomputed synthetic spectra of reflected light exoplanets, supernovae, galaxies, or even further afield such as laboratory physical chemistry, plasma physics, materials science, or remote sensing.

We place mild demands on the precomputed spectra. They should have sporadic regions of discernable continuum (or pseudocontinuum) devoid of lines, and the continuum shape should vary smoothly in wavelength. The spectral lines or pseudo-lines should be resolved (and not sub-sampled). Many precomputed synthetic stellar spectra will meet these criteria. For those spectra that do not meet these criteria, the method should still work with some additional fine-tuning of the pre-processing steps that follow.  M dwarfs in the red-optical, solar-like stars in the blue visible, and brown dwarfs at virtually any wavelength may fall into this category.

We truncate the red and blue limits of the precomputed synthetic spectrum to match a high-bandwidth echelle spectrograph, extended with a buffer at the edges of size $\pm \Delta \lambda_{\mathrm{buffer}}$.  The buffers are needed for two reasons: 1) to include the influence of spectral lines whose centers lie just outside the region-of-interest, but whose wings are broad enough to affect the neighboring pixels; and 2) to account for astrophysical radial velocity shifts that may send line centers into- and out of- the region of interest.  We set the value of the buffers to account for plausible radial velocity and rotational broadening of real stars. A generous buffer of $v \sin{i} < 500 \;\kmps$ and $|RV|<500 \;\kmps$ yields a typical buffer of about 30~\AA.

The choice of limiting the bandwidth to a region of interest around a single echelle spectrograph bandwidth stems from computational constraints. In principle, there is no fundamental limit to the bandwidth one could clone with the method presented here, up to and including the entire precomputed synthetic spectral model bandwidth. We adopt the exact native wavelength sampling with no smoothing or interpolation, yielding a wavelength vector $\bm{\lambda}_S$ with length $N_\mathrm{S}$ equal to the number of pixels within the extents of our region of interest including the buffers.

At this stage, we have the choice of whether to work in linear or log scale flux units. Adopting the log of the flux would ensure that the cloned model possesses only positive flux values, a desirable trait of any physical spectral model. We have implemented both modes in \texttt{blase}, allowing users to choose their preference.  We only narrate the linear flux unit description in the main text of this document for the sake of clarity, and since most practitioners may tend to think of flux in terms of linear flux units.  The data-model comparison step will always take place in linear flux units, so the only operational difference is the behavior for deep and saturated lines.  Appendix \ref{appendixLogScale} lists the equations adjusted to log flux units.

\subsection{Initialization}\label{subsecInit}

We initialize the cloned model with a series of preprocessing steps. We divide the entire spectrum by a black body $\mathsf{B}(\bm{\lambda}_S)$ of the same effective temperature $T_{\mathrm{eff}}$ as the model template. The resulting signal usually still has smooth wiggles around the continuum. An optional continuum flattening step ensures that subsequent spectral line finding steps get applied uniformly. This high-pass filtering step should be set to capture the genuine spectral shape, without over-fitting broad line wings such as those in deep hydrogen and sodium lines. Any high-pass filter will work, a Gaussian Process approach would be ideal \citep{czekala15}.  Instead, we apply a simple and familiar heuristic: fit a polynomial $\mathsf{P}(\bm{\lambda}_S)$ to a few continuum peaks and divide out the trend.

The result should be a flattened ``continuum-normalized'' spectrum familiar to practitioners in high-resolution spectroscopy, with the continuum level close to unity.  It is this spectrum that will serve as the centerpiece of subsequent training steps.  We, therefore, drop any subscript and simply refer to this flattened spectrum as $\mathsf{S}$:

\begin{eqnarray}
    \mathsf{S} = \mathsf{S}_{\rm abs}/\mathsf{B} / \mathsf{P}
    \label{eqnFlattening}
\end{eqnarray}

\noindent where the division indicates element-wise division of these arrays or ``vectors'' of flux values.  For elementwise-multiplication, we will adopt the $\odot$ operator symbol to distinguish that the multiplication occurs elementwise between two ``vectors'' of spectral fluxes, as opposed to a dot product (scalar output) or matrix multiplication (matrix output).  This style of vector multiplication and division is standard in numerical array manipulations such as those in \texttt{numpy}, \texttt{IDL}, \texttt{matlab}, \emph{etc.}

\begin{figure}[hbt!]
    \centering
    \includegraphics[width=0.45\textwidth]{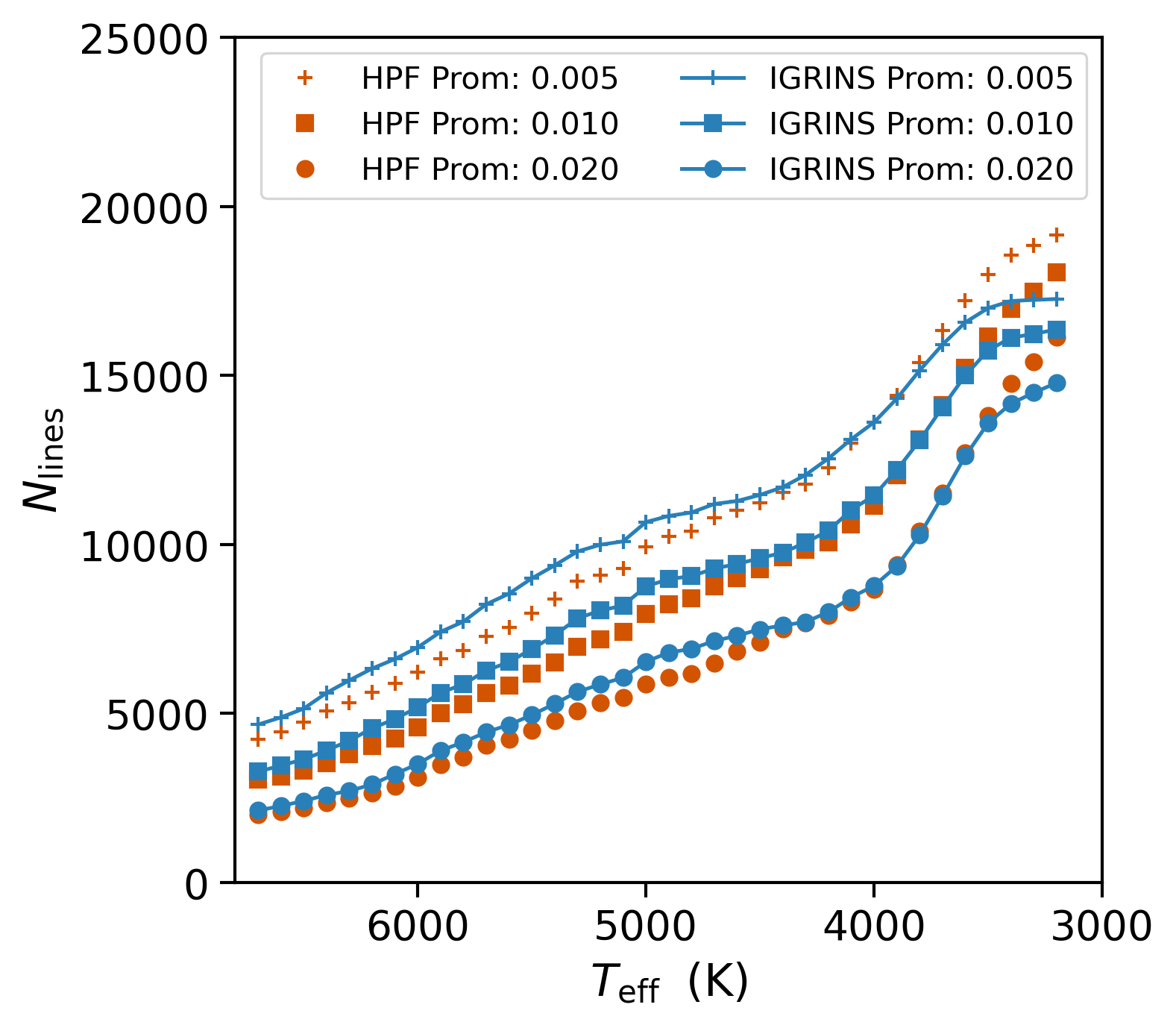}
    \includegraphics[width=0.41\textwidth]{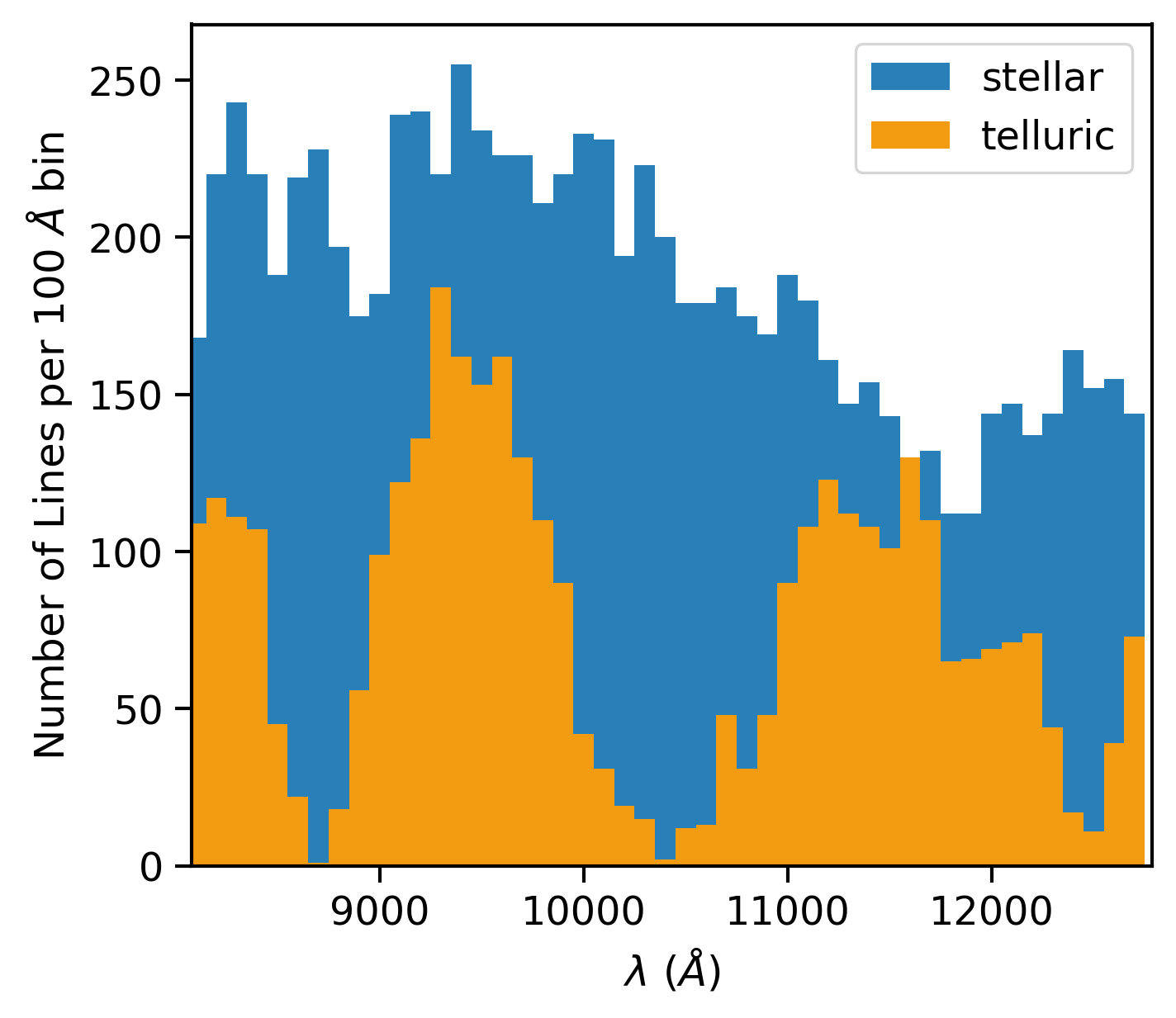}
    \caption{Scaling of prominence and density of spectral lines.
        \emph{Top:} Number of lines versus effective temperature for PHOENIX models truncated to IGRINS (\textbf{blue, connected} points) and HPF (\emph{red, free-standing} points) bandwidths, for different prominence thresholds of $0.02$, $0.01$, and $0.005$.
        \emph{Bottom:} Number of lines per 100 \AA\ wavelength bin for stellar (\emph{blue, upper envelope of steps}) and telluric (\emph{orange, lower envelope of steps}), illustrated for a $T_\mathrm{eff}=4700$~K, $\log{g}=4.5$ PHOENIX model and a $T=290 \mathrm{K}$, relative humidity of 40\% TelFit model.}
    \label{fig_Nlines_vs_teff}
\end{figure}

A recreation of the unvarnished input spectrum---if desired---can be obtained by multiplying the continuum-flattened signal by the ``perturbed black body'', $\mathsf{B}(\bm{\lambda}_S)\odot \mathsf{P}(\bm{\lambda}_S)$, that symbolizes the black body modulated by continuum opacity or broad-band radiative transfer effects.  This smooth spectrum may be useful for applications that need to keep track of broad-band flux, such as low-resolution spectra, or regions with molecular band heads. The ``perturbed black body'' continuum model contains $n_{\mathrm{poly}}+1$ fixed-but-possibly tunable lookup parameters, plus the fixed input $T_{\mathrm{eff}}$.  For most practitioners, these terms serve as nuisance parameters and are perfunctorily discarded.

Next, we identify the spectral lines. We apply the \texttt{find\_peaks} local-maximum-finding algorithm implemented in \texttt{SciPy} \citep{2020SciPy-NMeth} on the negative of the spectrum, in order to find the local minima.  We select a ``topographic prominence threshold'' $P_{\rm rom} \in (0.005, 0.02)$, defined as the vertical distance between the peak and the local baseline.  The baseline is defined with topographic heuristics that act as effective pseudo-continuum-detectors.  This threshold dictates the number of lines that will be modeled: a lower prominence finds more, weaker lines, and a larger prominence finds fewer, deeper lines. The prominence algorithm successfully finds lines that reside on top of broad line wings, or unresolved band heads provided that the individual lines exceed the prominence threshold in their local region. The number of lines $N_{\mathrm{lines}}$ depends on the bandwidth, prominence, and the intrinsic properties of the input spectrum, principally effective temperature and metallicity.

For this paper, we illustrate examples for two \'echelle spectrographs with particularly large spectral grasp: the Habitable Zone Planet Finder \citep[HPF,][]{2014SPIE.9147E..1GM} on the Hobby-Eberly Telescope at McDonald Observatory in Fort Davis, Texas; and the Immersion Grating Infrared Spectrograph \citep[IGRINS,][]{park14}
currently on the Gemini South Telescope on Cerro Pach\'on in Chile. The $R=55,000$ HPF has a native bandwidth of $8079-12785$~\AA, which we expand to $8049-12815$~\AA~including the edge buffers. IGRINS has two cameras for $H$ and $K$ band, with the combined spectrum spanning $14267-25217\;$\AA~ including the edge buffers and the region in-between the two cameras, all at a resolving power of $(R=45,000)$. The spectrograph acquisition, reduction, and post-processing steps yield data $\mathsf{D}(\bm{\lambda}_{D})$, where $\bm{\lambda}_{D}$ is the wavelength vector at the instrumental resolution and sampling of each instrument, generally much coarser than the resolution and sampling of the precomputed synthetic spectra. The data wavelength vector may also contain gaps between \'echelle orders, whereas the precomputed wavelength coordinates are usually contiguous. HPF may have up to $2048\times28=$57,344 pixels, and IGRINS has typically about 75,000 pixels, after common trimming of noisy edge pixels and unusable telluric regions.  Meanwhile, the HPF-truncated model spectra have $N_s=$335,849 native resolution samples, comparable to the IGRINS-truncated model spectra, $N_s=$330,052.

Figure \ref{fig_Nlines_vs_teff} shows how the number of detected lines $N_{\mathrm{lines}}$ scales with effective temperature and prominence threshold $P_{\rm rom}$ for the \texttt{PHOENIX} grid, truncated to the bandwidths-plus-buffers for HPF and IGRINS. We see between about 2,000 and 20,000 lines depending on the $T_{\mathrm{eff}}$ and $P_{\rm rom}$. HPF and IGRINS have a comparable number of lines, and halving the prominence increases the number of lines by about $20-30\%$ in these ranges. The number of lines monotonically increases towards cooler effective temperatures.

So far we have only one piece of information about the spectral lines: their location. Next, we derive coarse properties about each detected peak: its amplitude and width, again using the prominence algorithms implemented in \texttt{scipy} \citep{2020SciPy-NMeth}.

There does not exist a general-purpose, single-shot algorithm for obtaining the lineshape in the presence of overlapping spectral lines: where do the wings of one line begin and the wings of another adjacent line end? We, therefore, do not attempt to determine anything about the lineshape at this stage and instead assume that the lines resemble a Voigt profile, with a guess width about equally split between Lorentzian and Gaussian.

\subsection{The \emph{blas\'e} Stellar Clone Model}

We have now arrived at the \emph{blas\'e} clone model $\mathsf{S}_{\rm clone}(\bm{\lambda}_S)$ for a flattened synthetic spectrum $\mathsf{S}$: it is the cumulative product of transmission through the sea of all overlapping spectral lines:

\begin{eqnarray}
    \mathsf{S}_{\rm clone} = {\displaystyle \prod_{j=1}^{N_{\mathrm{lines}}} 1-a_j \mathsf{V}_j } \label{equation1}
\end{eqnarray}

\noindent where $\mathsf{V}_j$ is the Voigt profile $\mathsf{V}(\bm{\lambda}_S-\lambda_{\mathrm{c},j}, \sigma_j, \gamma_j)$ with Gaussian standard deviation $\sigma$, Lorentzian half-width $\gamma$, at line center position $\lambda_c$, for the $j^{th}$ spectral line. The amplitude $a$ is always expected to be positive for absorption lines.

The Voigt profile $\mathsf{V}(\lambda, \sigma_j, \gamma_j)$ can be computed in exact closed-form using the Voigt-Hjerting function \citep{1938ApJ....88..508H} as the real part of the complex Fadeeva function \citep[\emph{e.g.}][]{2011arXiv1106.0151Z}. Evaluation of the Fadeeva function can be computationally costly, so approximate forms may be desirable. Here we adopt the pseudo-Voigt approximation \citep{Ida:nt0146}.

\begin{figure*}[hbt!]
    \centering
    \includegraphics[width=1.0\textwidth]{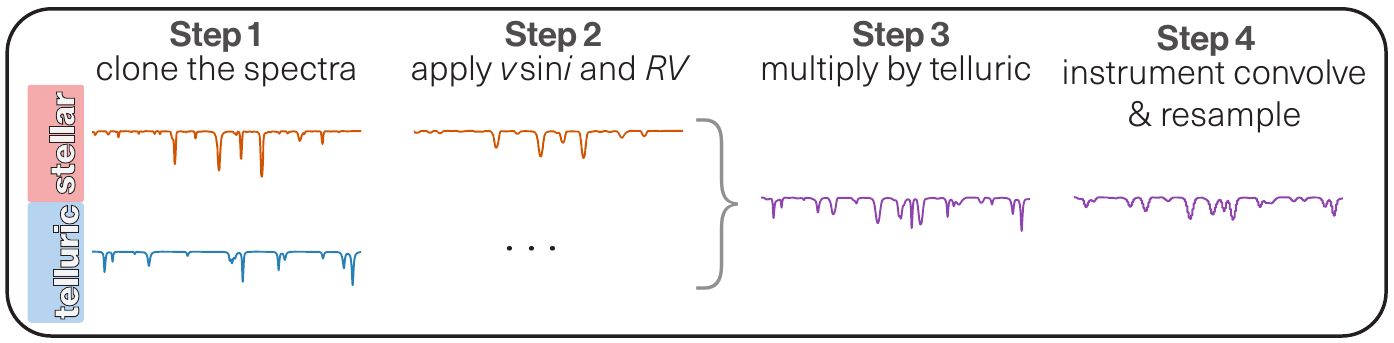}
    \caption{Visual flowchart of the  \emph{blas\'e} forward model.  \textbf{Step 0} (not shown) is to choose a precomputed synthetic stellar spectrum---and, optionally, a precomputed synthetic telluric spectrum---with physical properties close to the target and observing conditions.  Both the stellar and telluric spectra get cloned (\textbf{Step 1}).  The stellar model is warped to its extrinsic properties (\textbf{Step 2}), and then the stellar and telluric models get multiplied together (\textbf{Step 3}).  Next, this joint model is convolved with an instrumental kernel and resampled to the wavelength coordinates of the data spectrum (\textbf{Step 4}).  It is this forward model that gets directly compared to the observed spectrum (not shown).
    }
    \label{blase_flowchart}
\end{figure*}

\subsection{Goodness of fit metric and gradients}
The model evaluated with its coarse initial values would have terrible performance: it would only vaguely resemble the synthetic spectral model, with up to $\pm 50\%$ undulations from the inexact assignment of widths, lineshapes, and amplitudes. Instead, we tune the parameters of the model, starting from these coarse initial values. This model has between $N_{\mathrm{lines}}\times 1$ and $N_{\mathrm{lines}}\times 4$ free parameters, depending on how many of the 4 physical line properties you wish to tune.  The threshold-based line-finding algorithm performs very well at finding the line center positions, so in practice, we can fix the line center wavelengths.  We, therefore, default to fitting three parameters per line, where the center wavelength is held fixed and the amplitude, width, and lineshape are allowed to vary. We minimize a scalar ``goodness-of-fit'' metric, \emph{aka} loss scalar $\mathcal{L}$, chosen as the mean squared error (MSE), which is proportional to $\chi^2$, the sum of the squares of the residual vector $\mathsf{R} \equiv \mathsf{S}-\mathsf{S}_{\rm clone}$ but has no notion of per-pixel noise since the precomputed synthetic spectrum has no uncertainty:

\begin{eqnarray}
    \mathcal{L} = \sum_i^{N_S} (S_i - S_{\mathrm{clone},i})^2 = \mathsf{R^\intercal}\cdot \mathsf{R} \label{simpleLikelihood}
\end{eqnarray}

As seen in Figure \ref{fig_Nlines_vs_teff}, the number of lines can exceed 7,000, meaning the clone model has over 21,000 free parameters. Fitting more than about 300 parameters is difficult with conventional optimizers, and so historically the practice of finding the best settings for all 21,000 parameters would have been considered hopeless, a fool's errand.  A new category of ``gradient-based'' optimizers offers a breakthrough.  Gradient-based optimizers have virtually no restriction on the number of parameters, scaling to functions of possibly millions or billions of tuning parameters.  Gradient-based optimizers get vastly more information per likelihood evaluation: $N_\mathrm{params}+1$ pieces of information-- the scalar value of the loss evaluated for some setting, \emph{and} the derivative of the loss scalar with respect to each of the parameters, the so-called \emph{Jacobian}: $\grad \mathcal{L} = (\frac{\partial \mathcal{L}}{\partial a_j}, \frac{\partial \mathcal{L}}{\partial \sigma_j}, \cdots, \frac{\partial \mathcal{L}}{\partial \gamma_j})$.  The Jacobian indicates how the MSE would decrease with a change in the parameter-of-interest, or put simply ``which way and by how much'' you have to change each individual line property to get a better fit.

Geometrically, the $\grad \mathcal{L}$ Jacobian resembles a compass needle guiding the direction to the best fit. This compass needle resides in, say, 21,000-dimensional space, instead of our mere 3 dimensions of physical space.  Conventional gradient-free optimizers evaluate the Jacobian through approximate finite differences, which result in numerical instabilities.  Those instabilities accrue and break down at around the 300-dimension mark.  So non-gradient-based optimization can be pictured as having only a coarse needle---assembled from approximate finite differences---that begins to spin aimlessly after too many dimensions.

We adopt ``Full-batch'' Gradient Descent optimization, as opposed to the more popular Stochastic Gradient Descent \citep[SGD,][]{2016arXiv160904747R}.  In Full-Batch Gradient Descent, the entire dataset is evaluated in Equation \ref{simpleLikelihood}, as opposed to only a portion of the dataset in the Stochastic counterpart.  Appendix \ref{minibatches} discusses the tradeoffs and rationale for ``Full-Batch''.

The optimizer updates the $a_j, \sigma_j, \gamma_j$ parameters by a small fraction of the Jacobian---called the learning rate (LR)---towards the direction that would improve the fit, for all parameters \emph{simultaneously}. The Jacobian is calculated behind the scenes with automatic differentiation implemented as the so-called backpropagation algorithm or simply ``backprop'' \citep{2015arXiv150205767G}. We choose the \texttt{PyTorch} framework that computes these Jacobians efficiently for all of the mathematical primitives in our \texttt{blase} implementation \citep{2019arXiv191201703P}.

It is this ability to automatically compute Jacobians that sets \texttt{PyTorch} (and \texttt{JAX} and \texttt{TensorFlow}) apart.  These frameworks give exact Jacobians, instantaneously, for free (or cheap).  Without exact Jacobians, a conventional optimization step only obtains \emph{one} piece of information: how much the overall loss changed.  With exact Jacobians, we obtain $N_\mathrm{params}+1$ pieces of information for each evaluation of the forward model.  The power of gradient descent becomes transformative as the number of parameters grows into the tens of thousands, as is the case for \emph{blas\'e}.

\subsection{GPU and Autodiff specific considerations}\label{gpuConsiderations}

Forward modeling with tens of thousands of physics-informed parameters may seem like such a significant paradigm shift that it can feel too good to be true.  In this section, we introduce the non-negligible architectural design tradeoffs that arise when adopting an autodiff framework---such as \texttt{PyTorch}---for physics-based forward modeling.

First, we make a few tweaks to the implementation for numerical purposes. We enforce that all Gaussian and Lorentzian widths are positive by tuning the natural log of the widths, then exponentiating them before inclusion in Equation \ref{equation1}. Based on the minus sign in Equation \ref{equation1}, an amplitude $a_j$ could hypothetically take on either positive values (flux loss, absorption) or negative values (flux gain, emission).  For now, we focus on photospheric absorption lines---as opposed to, say, chromospheric emission lines---and therefore enforce all the amplitudes to be positive by tuning $\ln{a_j}$ and then exponentiating in the same strategy as above.  Emission lines---if desired---could be included with a mere sign flip to Equation \ref{equation1}.  Lines that can manifest either in absorption or emission could hypothetically relax the natural log pre-processing step for isolated lines.  We narrate only the absorption scenario moving forward.

The autodiff machinery has a convenient way to set which parameters are held fixed and which are iteratively fine-tuned.  One simply disables the autodiff flag for the fixed parameters: we set the \texttt{requires\_grad=True} property for any \texttt{PyTorch} tensor that we want to vary. This design allows us to easily explore whether, say, allowing the $\lambda_\mathrm{c}$ parameter to vary significantly improves the fit.

The computational bottleneck occurs at the evaluation of Equation \ref{equation1}, which can be viewed as having an $N_{\mathrm{lines}}\times N_{S}$ matrix $\bm{\bar{F}}$ assembled by stacking each Voigt absoption profile $\mathsf{V}_j(\bm{\lambda}_s)$ on top of each other:

\begin{equation}
    \begin{pmatrix}
        1 - a_1 \mathsf{V}_1(\bm{\lambda}_s)                                       & \\
        1 - a_2 \mathsf{V}_2(\bm{\lambda}_s)                                       & \\
        \vdots                                                                     & \\
        1 - a_{N_{\mathrm{lines}}} \mathsf{V}_{N_{\mathrm{lines}}}(\bm{\lambda}_s) &
    \end{pmatrix}
\end{equation}

An element of this matrix, $F_{ji}$, will have the flux value for a given $j^{th}$ line at a given $i^{th}$ wavelength coordinate. Equation \ref{equation1} performs a type of matrix contraction, turning an $N_{\mathrm{lines}}\times N_{S}$ matrix into a length $N_{S}$ row vector. The number of Floating Point Operations (FLOPS) scales with the number of entries in this matrix. So we face a tradeoff of wanting to make the matrix large for accuracy and small for computational expedience.

We can rewrite Equation \ref{equation1} as a sum by taking the log of both sides and dropping in this $\bm{\bar{F}}$ matrix:

\begin{eqnarray}
    \ln{\mathsf{S}_{\rm clone}} = \sum_{j=1}^{N_{lines}} \ln{F_{ji}} = \mathbf{1} \cdot \ln{\bm{\bar{F}}}  \label{eqnFbar}
\end{eqnarray}

\noindent where $\mathbf{1}$ is a $1\times N_{\rm lines}$ row vector of all-ones. We re\"emphasize that---in its current form---each spectral line has to be painstakingly evaluated across the entire spectral bandwidth.   Efficient GPU algorithms exist for voluminous matrix manipulations such as this one, so this voluminous computation will proceed as quickly as possible on modern machines. In particular, the proprietary CUDA architecture for NVIDIA\textsuperscript{\tiny\textregistered} GPUs contains Tensor cores with specialized matrix math. The chief bottleneck occurs when the storage of the $\bm{\bar{F}}$ matrix exceeds the available RAM of a GPU or CPU: the computation will fail with an ``Out of Memory'' exception. Modern NVIDIA GPUs have $8-40$ GB of RAM, which translates roughly to a few thousand spectral lines across $\sim$300,000 pixels.  It is generally not possible to construct Equation \ref{eqnFbar} in its entirety in one-fell-swoop, even on a GPU. A remedy is needed.

\subsection{Sparsity}

The $\ln{\bm{\bar{F}}}$ matrix is sparsely populated: most of the entries far from the line center are vanishingly close to zero. Here we take advantage of that mostly empty matrix using the mathematics of sparse matrices \citep{saad03:IMS}.

We retain a relatively small number of pixels $N_{\rm cut}$ adjacent to the line center. Setting this wing cut produces a speedup by a factor of $\frac{N_S}{N_{\mathrm{cut}}}$, which can exceed $100\times$ for wide bandwidth spectra. The choice of $N_{\rm cut}$ is nuanced.  It should be set large enough that truncation effects are not seen for the broadest lines.  But even more, $N_{\rm cut}$ has to be future-proofed for Doppler-shifting. Extreme Doppler shifts could hypothetically send line centers entirely outside the extents of $N_{\rm cut}$ if set too low.  We therefore typically set wingcuts comparable to the buffer size $2 \Delta \lambda_{\mathrm{buffer}}$, even though most weak lines only perceptibly affect $<1\;$\AA. We coerce all wing cuts to be the same number of pixels, typically 6000 pixels, $\sim30-60~$\AA~ for \texttt{PHOENIX}, with the middle pixel being at the line center position, and about 3000 pixels to the red and blue side of the line. We populate a new approximate sparse matrix $\ln{\bm{\hat{F}}}$ with only these $6000$ pixels per line and assume zeros everywhere else.

The remapping of the sparse matrix can be pictured as having shifted all lines to the center of this new matrix $\bm{\hat{F}}$, visualized pictorially in Equation \ref{eqnPictograph}.  The algorithmic machinery keeps track of each $(i, j, F_{ji})$ trio of coordinates and flux values.

\begin{equation} \label{eqnPictograph}
    \begin{aligned}
        \bm{\bar{F}} & =
        \begin{pmatrix}
            \cdots \includegraphics[height=1cm]{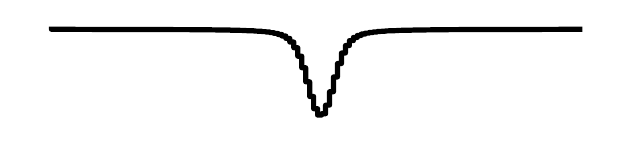}  \cdots & \\
            \cdots \includegraphics[height=1cm]{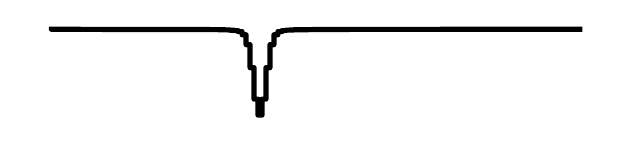}  \cdots & \\
            \vdots                                                    & \\
            \cdots \includegraphics[height=1cm]{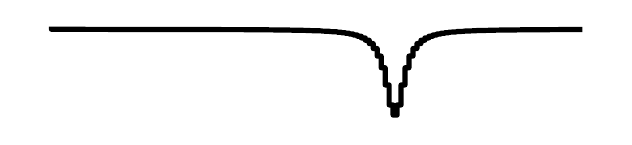} \cdots  &
        \end{pmatrix}                        \\
        \bm{\bar{F}} & \mapsto \bm{\hat{F}} =\begin{pmatrix}
                                                 \includegraphics[height=1cm]{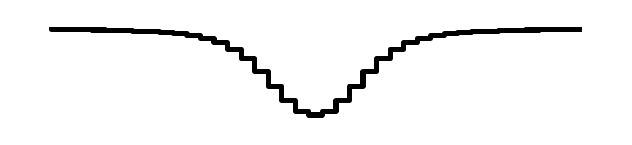} & \\
                                                 \includegraphics[height=1cm]{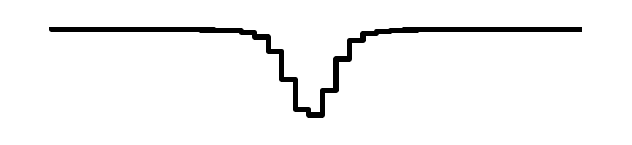} & \\
                                                 \vdots                                      & \\
                                                 \includegraphics[height=1cm]{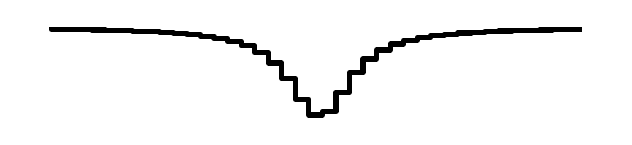} &
                                             \end{pmatrix}
    \end{aligned}
\end{equation}

Sparse matrix methods generally support an operation known as \emph{coalescing}, which sums values with repeated indices. Each pixel may get computed about $\sim100$ times in this sparse implementation, which is about $70\times$ faster than each pixel getting computed $N_\mathrm{lines}\sim7000$ times in the dense approach.  Efficient algorithms for assembling and coalescing sparse matrices exist in \texttt{PyTorch}.  Some GPUs now support additional hardware acceleration of sparse matrices, providing even greater speedups.

\subsection{Optimization and training}

We use the Adam optimizer \citep{2014arXiv1412.6980K} with a typical learning rate $LR\in (0.005, 0.1)$ and all the defaults for \texttt{PyTorch v1.11}.  We defined the number of training epochs $N_{epoch}=100-10,000$ depending on the application. The user can optionally monitor a live view of the training progress with Tensorboard \citep{tensorflow2015-whitepaper} to gain an intuition for the training efficiency.

\begin{figure}[hbt!]
    \centering
    \includegraphics[width=1.0\columnwidth]{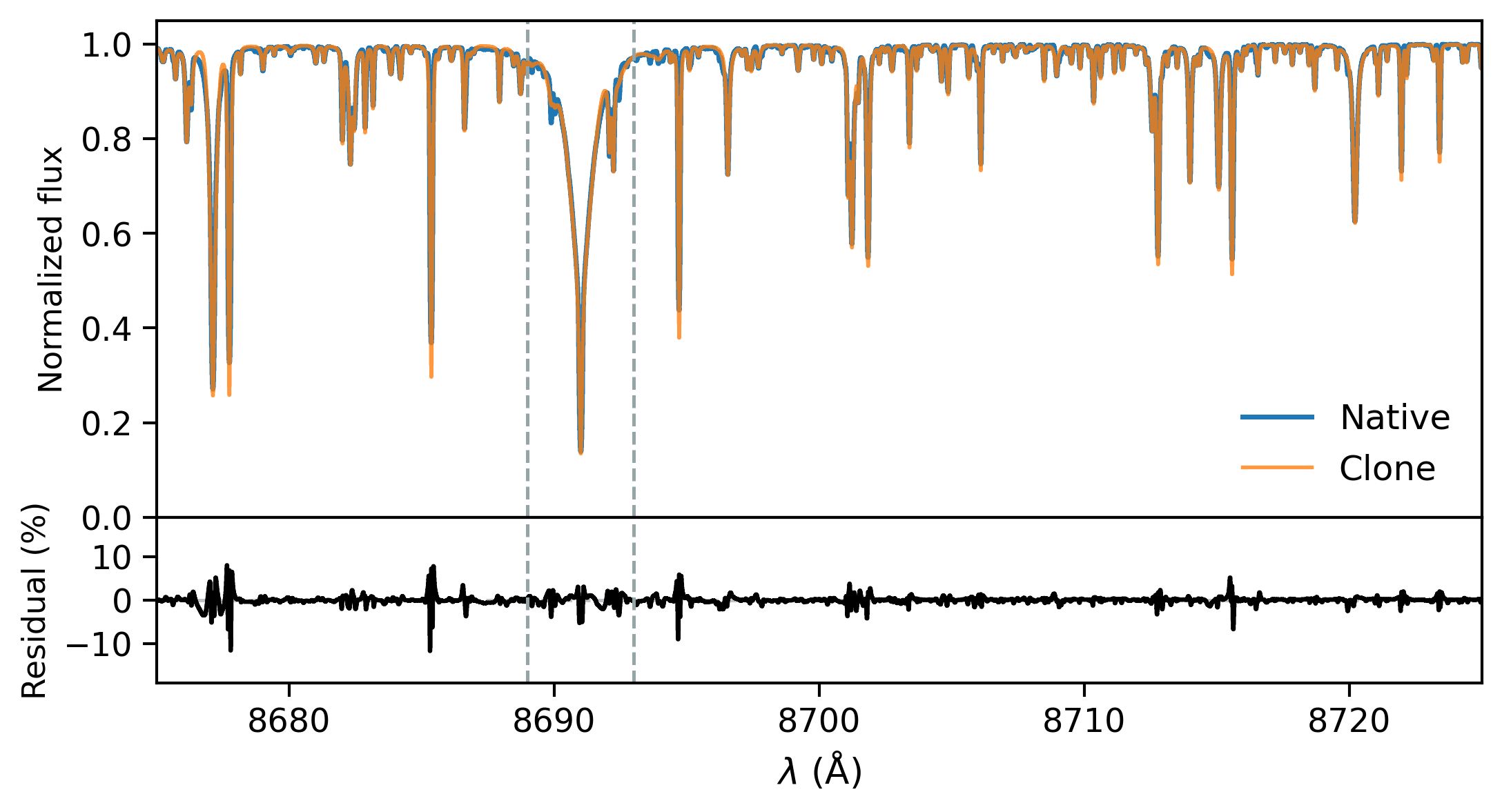}
    \caption{PHOENIX spectrum cloned with \emph{blas\'e}.  This $T_{\mathrm{eff}}=4700\;$K,  $\log{g}=4.5$ solar metallicity model has 9,028 individual cloned spectral lines, each with 3 tuned parameters.  The pictured 50 \AA~ chunk contains 121 spectral lines and represents about 1$\%$ of the entire spectral bandwidth that was cloned.  Some flaws can be seen near the cores of deep lines or wings of broad lines.}
    \label{fig_cloned_spectrum_demo}
\end{figure}

Figure \ref{fig_cloned_spectrum_demo} shows a portion of a PHOENIX spectrum cloned with \texttt{blase}. The $1000$ epochs of training took 56 seconds on an NVIDIA\textsuperscript{\tiny\textregistered} RTX2070 GPU with \texttt{PyTorch v1.11}, \texttt{CUDA v11.1}, and Intel\textsuperscript{\tiny\textregistered} Core\textsuperscript{\tiny TM} i7-9750H CPUs at 2.60GHz, with all tensors as FP64. The same computation on a 2020 M1 Macbook Air took 1$^h$25$^m$ with \texttt{PyTorch v1.9}, $90\times$ slower than the GPU counterpart.

We store the model parameters to disk and refer to the entire collection of parameters as a \emph{pre-trained model}.  More specifically this fine-tuned model represents an evaluable and interpretable clone of the original static pixel-by-pixel flux values.

\section{Methodology II: Cloning Telluric Spectra} \label{sectionTelluric}
Ground-based near-IR \'echelle spectra possess thousands of depressions attributable to molecular line absorption in Earth's atmosphere.  These telluric lines hamper the unbiased interpretation of \'echelle spectra, so some treatment plan is needed.  Often the regions of known, deep tellurics are simply discarded.  In other cases, the lines are modeled with first principles line-by-line radiative transfer \citep[\emph{e.g.} \texttt{TelFit,}][]{2014AJ....148...53G, 2005JQSRT..91..233C} or through data-driven means \citep[\emph{e.g.} \texttt{wobble,}][]{2019AJ....158..164B}.  The most demanding EPRV applications require a precision characterization of telluric lines that the astronomical community has not yet been able to achieve, and that may rival even the abilities of Earth Science practitioners.  A hybrid data-/model- driven approach was among the chief recommendations of the \emph{Telluric Hack Week} Workshop\footnote{\url{https://speakerdeck.com/dwhgg/telluric-line-hack-week-wrap-up}} aimed at improving mitigation of the atmosphere's deleterious effects.  The  \emph{blas\'e} framework achieves a key milestone by introducing a \emph{hybrid} approach to tellurics.

\subsection{The  \emph{blas\'e} Telluric Clone Model}
We start with a precomputed synthetic telluric model, $\mathsf{T}$ with associated wavelength coordinates $\bm{\lambda}_T$.  We employ a \texttt{TelFit} model, though any precomputed synthetic telluric model will work, such as \texttt{MOLECFIT} \citep{2015A&A...576A..77S}.  The \texttt{TelFit} model does not contain any continuum sources of opacity, so we can skip the flattening procedure described in Equation \ref{eqnFlattening}.  We orchestrate the same initialization and line finding as in the stellar models and obtain a coarse clone.

The number of pixels in the telluric model can be chosen at the time of running \texttt{TelFit}.  Here we choose a spectral resolution $R\sim10^6$, adequate for resolving narrow telluric lines, and yielding about 2 million pixels across the entire HPF bandwidth.  This pixel sampling is about $6\times$ finer than the native PHOENIX pixel sampling.  The number of telluric lines depends on the atmospheric properties, in particular, the local surface temperature $T_\oplus$ and relative humidity $RH$.  For a surface temperature of $T=290 \mathrm{K}\; (62^\circ~\mathrm{F})$, relative humidity of 40\%, and typical conditions for McDonald Observatory, we anticipate 3615 telluric lines across the entire HPF bandwidth, distributed as shown in the bottom panel of Figure \ref{fig_Nlines_vs_teff}.

One guiding principle departs from the stellar case: telluric lines do not require future-proofing for large radial velocity shifts, hypothetically allowing us to reduce the number of pixels needed for a wingcut.  Small radial velocity shifts are possible due to bulk motions in the Earth's atmosphere, but those bulk motions should be much smaller than the speeds of stars towards and away from Earth.  So we can hypothetically tolerate a much smaller wing cut for telluric lines.  In practice, telluric lines can be saturated, and accurately cloning the resulting broad telluric wings still benefits from $N_{\rm cut}\sim6000$ pixels, comparable to the stellar scenario.

\begin{figure*}[hbt!]
    \centering
    \includegraphics[width=0.98\textwidth]{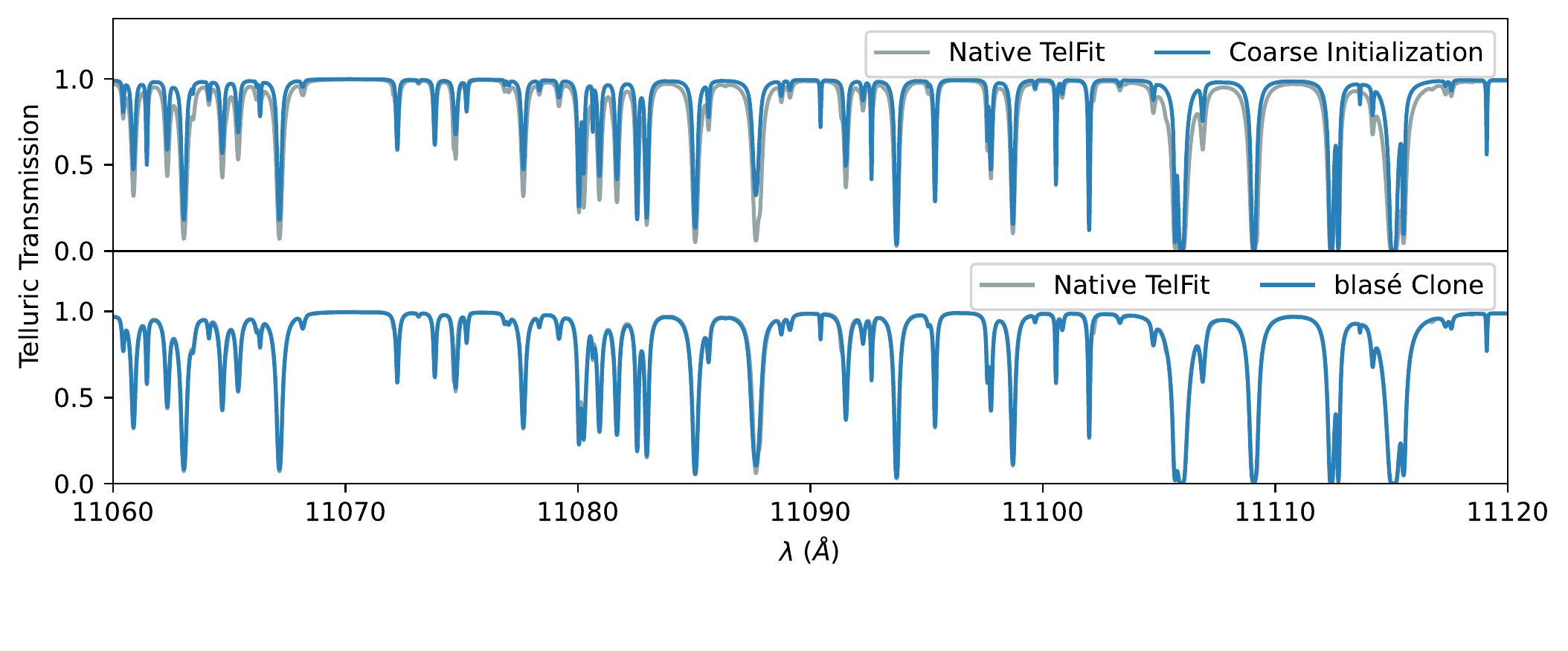}
    \caption{Pre-computed TelFit model cloned with \texttt{blase}.  \textbf{Before:} The top panel shows a forward model evaluated with coarsely initialized line-by-line properties, based on simple threshold-based peak-finding and crude estimation of the shape of the line.  \textbf{After:} The bottom panel shows the \texttt{blase} fine-tuned clone, with nearly identical spectral structure.}
    \label{fig_telluric_clone}
\end{figure*}

We optimize the sparse telluric clone, achieving comparable computational speed as the PHOENIX cloning task.  We are left with $\mathsf{T}_{\rm clone}(\bm{\lambda}_T)$, the tunable telluric clone model evaluated at its original native coordinates.  Figure \ref{fig_telluric_clone} shows a before-and-after view of the  \emph{blase} cloning procedure for a TelFit model zoomed in on a $J-$band region with dense Telluric lines.  We see that the blas\'e fine-tuning appears nearly indistinguishable from the native resolution TelFit precomputed model.

\section{Cloning Performance}

We compute the residual $\mathsf{R}(\bm{\lambda}_S)$ of native-minus-cloned PHOENIX model, illustrated in the bottom panel of Figure \ref{fig_cloned_spectrum_demo}. We see an RMS residual of 1.2\%/pixel at native resolution.  The telluric clone shows a comparable level of performance.  These residuals tend to pile up in local symmetrically balanced clusters that get canceled out once convolved with coarser instrumental line profiles, so their overall effect at instrumental resolution is typically negligible: the clones are almost perfect.  We identify three main categories of cloning flaws that may not be negligible depending on the science application.

The first---and expected---source of large residuals is simply missing line opacity due to our finite prominence threshold. Lines with prominence less than $P_{\rm rom}$ yield residual notches with strengths comparable to $P_{\rm rom}$. Including smaller prominence lines by lowering $P_{\rm rom}$ produces smaller residuals, at the tradeoff of computing more lines and yielding higher computational cost.  But at some point, turning down $P_{\rm rom}$ yields diminishing returns, as other imperfections provide a noise floor.  We have experimentally determined this noise floor to occur near $P_{\rm rom}=0.01$.

Second, a conspicuous flaw occurs in the line cores of relatively narrow lines.  The cloned model tends to overestimate the flux at the core and underestimate the flux along the slopes of the lines.  The residuals, therefore, exhibit a \texttt{W}-shape ringing artifact.  The scientific impact of this flaw will depend on the precision demands of the application, but it may not be as bad as it seems for many applications.  Recall that the cloning occurs at native spectral resolution, much greater than typical instrumental resolution.  The \texttt{W}-shape ringing residuals will tend to cancel each other out once the clone model undergoes convolution with an instrumental profile (Section \ref{jointModel}).  We ascribe this flaw to the adoption of approximate pseudo-Voigt profiles rather than exact Voigt profiles.  High-precision applications may wish to employ the exact Voigt profile, which we estimate to cost about 27$\times$ more than this approximate form.

The overprediction of line cores in narrow lines is only partly attributable to the pseudo-Voigt approximation.  Even exact Voigt profiles inadequately represent true atmospheric opacities. Typical stellar and substellar atmosphere models compute opacities for a stratified atmosphere consisting of tens or hundreds of locally isothermal, isobaric atmospheric layers.  Each of those layers may exhibit a local per-line opacity that matches a Voigt profile.  But the cumulative product of many slightly-different Voigt profiles is not exactly equal to any single Voigt profile.  So the \emph{blas\'e} approach of asserting lines as Voigt profiles can be best understood as a phenomenological model-- that lines tend to look Voigt-like, even if they are non-Voigt in detail.  In principle, \texttt{blas\'e} could be configured in a way to learn the requisite perturbations to the Voigt profile in a systematic manner, pooling information among lines.  Such an approach is not yet implemented and may be an enticing avenue for future research.

\begin{figure}[hbt!]
    \centering
    \includegraphics[width=0.48\textwidth]{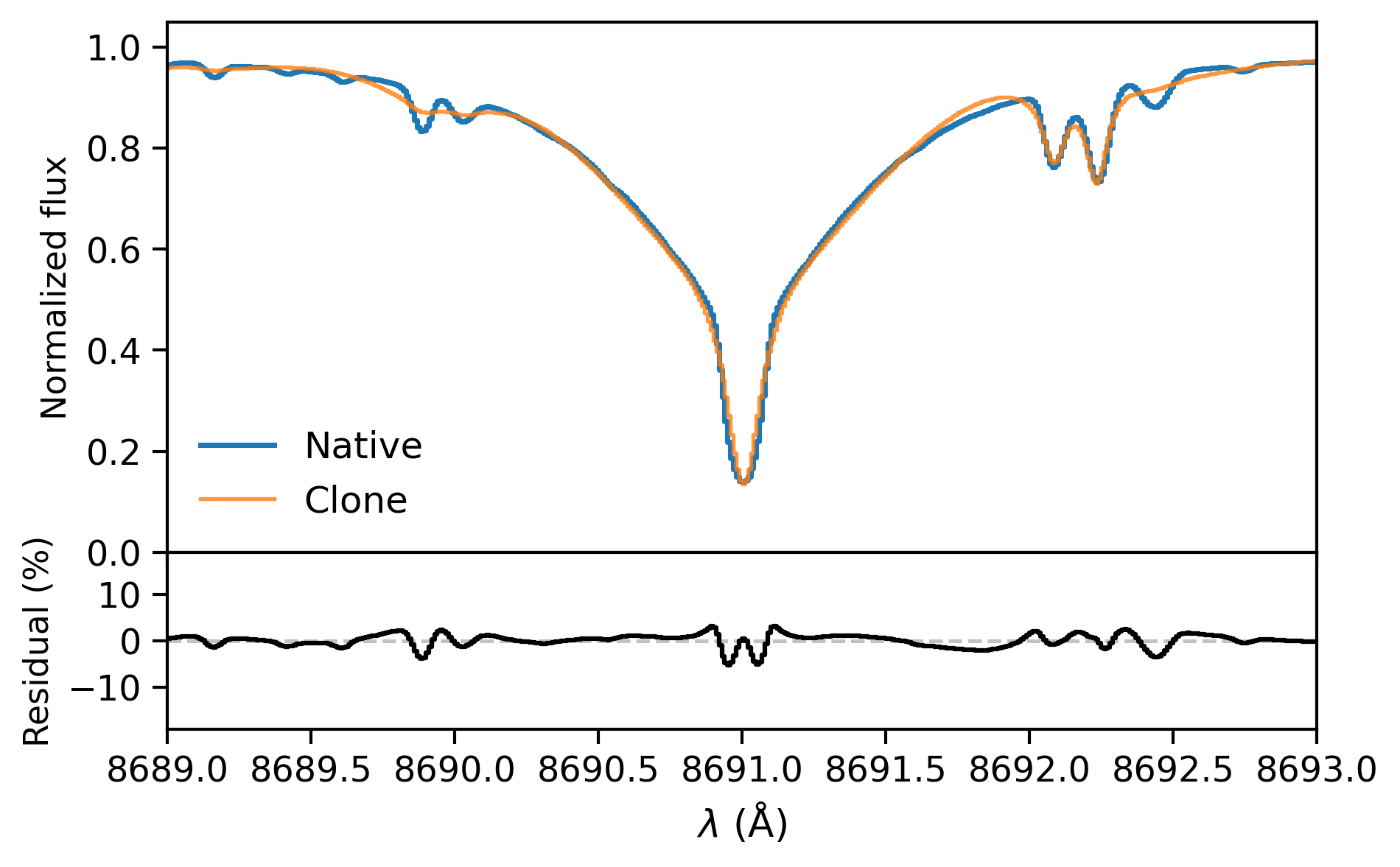}
    \caption{Zoom in of the region between the vertical gray bars in Figure \ref{fig_cloned_spectrum_demo}. The cloned model has 7 spectral lines describing the 400 pixels in this 4 \AA~ chunk.  The native PHOENIX pixel sampling can be seen as the boxy steps in both the native and cloned model.  }
    \label{fig_zoom_cloning_performance}
\end{figure}

Finally, and most perniciously, a large category of residuals appear near the wings of the deepest and broadest lines---such as hydrogen lines and neutral alkali metal lines.  The true lines exhibit advanced lineshapes, such as non-Lorentzian line wings that are not captured with the overly simplistic Voigt line profile, even an exact one.  Figure \ref{fig_zoom_cloning_performance} highlights super-Lorentzian line wings around a line at 8691~\AA.  Nearby narrow lines devolve into missing line wing opacity, the favored tradeoff when the continuum estimate's poor performance outweighs the pain of a narrow-but-tolerably-small spike. This flaw can be seen where a line initialized at 8692.5~\AA~ and another pair of lines at 8690.0 all melt into line wings.

Cloning telluric lines suffers from one additional problem.  Telluric lines can be extremely deep, exhibiting almost vanishing transmission with saturated line cores common in-between the atmospheric windows that define the $I$, $J$, $H$, and $K$ bands.  The \texttt{blas\'e} method can cope with these saturated lines, but often it treats nearby-and-blended saturated lines as one single line.  This glomming together of lines has little practical effect since few astronomical practitioners can make use of such profoundly saturated data.

\section{Semi-empirical models with transfer learning techniques}\label{transferLearn}

The cloned model already represents a useful intermediate product: the distillation of $N_s=335,849\times2$ pixel flux values and their wavelength coordinates into a more compact quartet of properties for a list of $<10,000$ spectral lines, a dimensionality reduction of $18\times$ for the cost of 1.2\%/pixel in accuracy.  So the process so far can be myopically viewed as a physics-informed compression algorithm.  But the cloned model serves as a mere stepping stone in our principal quest: the comparison of models to real data.

\subsection{Augmenting the stellar clone with radial velocity and rotational broadening }
Real stars possess two key extrinsic properties.  Rotational broadening $v\sin{i}$ and radial velocity $RV$ depend on the observer's viewing location. We follow \citet{czekala15} by emphasizing the qualifier \emph{extrinsic}, to distinguish between stellar \emph{intrinsic} properties, such as $T_{\mathrm{eff}}, \log{g},\mathrm{and\,} [\mathrm{Fe}/\mathrm{H}]$.  Intrinsic properties appear the same from any viewing location---at least for stars with isotropic surfaces---while extrinsic properties do not.  The distinction is important because the extrinsic terms act as simple convolutions and translations to the cloned spectrum, and can be treated after the cloning procedure.  We, therefore, define an augmented model, which we designate the ``extrinsic'' model, $\mathsf{S}_{\rm ext}$:

\begin{eqnarray}
    \mathsf{S}_{\rm ext}(\bm{\lambda}_Z) = \mathsf{S}_{\rm clone}(\bm{\lambda}_Z - \frac{RV}{c}\lambda_\mathrm{c}) * \zeta \left(\frac{\bm{v}}{v\sin{i}}\right) \label{eqn_convolution}
\end{eqnarray}

\noindent where $\zeta$ is the convolution kernel for rigid body rotation \citep[\emph{e.g.}][]{2022ApJS..258...31K}, $\bm{v}$ is the spectral axis represented as relative velocity coordinates, and $*$ denotes the convolution operator.

Most autodiff-aware \texttt{convolve} operators act in pixel-space, approximating kernels as numerically sampled functions.  There exists a special exponential spectral sample spacing that allows the convolution operators to work out of the box for rotational broadening.  The design of \emph{blas\'e} permits the stellar and telluric models to be reevaluated at any wavelength coordinate vector, provided that it adequately samples the underlying lines.  We, therefore, change the sampling from the native stellar and telluric wavelength coordinate grids, $\bm{\lambda}_S$ and $\bm{\lambda}_T$, to this special exponentially sampled wavelength grid, denoted with the subscript $Z$:

\begin{eqnarray}
    \bm{\lambda}_Z = \lambda_0  \exp{\frac{\bm{v}-v_0}{c}}
\end{eqnarray}

\noindent where $\bm{v}-v_0$ is the velocity vector going from zero to the velocity associated with the largest wavelength, with \emph{linear} spaced velocity samples.  We choose a sampling in velocity space of 0.5 km/s, which corresponds to about $10\times$ finer than the instrumental resolving power of HPF, and delivers a minimum and maximum wavelength spacing of 0.013 and 0.024 \AA$/\text{pixel}$ respectively for the HPF bandwidth.

Operationally, the radial velocity shift $RV$ gets applied to the line center positions rather than scaling the entire wavelength grid point coordinates, $\bm{\lambda}_Z$. This choice yields a convenience: it cleanly makes the $RV$ autodiff-aware, meaning that an infinitesimal change to the RV value can be sensed through backpropagation by affecting only the line center positions.

The $\mathsf{S}_{\rm ext}$ spectrum is shown in Step 2 of Figure \ref{blase_flowchart}.  There is currently no equivalent post-processing of the telluric spectrum $\mathsf{T}$.  As mentioned previously, we assume the motions of Earth's atmosphere are much less than the desired stellar radial velocity precisions.  However, demanding Extreme Precision Radial Velocity (EPRV) exoplanet searches may need to consider minuscule systematic $RV$ shifts and broadening of the telluric templates, arising from the turbulent and bulk motions of the Earth's atmosphere.  A $\mathsf{T}_{\rm ext}$ could hypothetically be included in  \emph{blas\'e} to achieve these strenuous precision demands.

\subsection{Joint Stellar and Telluric Model}\label{jointModel}
Figure \ref{blase_flowchart} shows a visual guide to all the steps in  \emph{blas\'e}.  We have arrived at what may be the most intriguing-and-yet-simple of these steps: we simply multiply the rotationally-broadened-and-$RV$-shifted stellar model by the telluric transmission:

\begin{eqnarray}
    \mathsf{M}_{\rm joint} = \mathsf{S}_{\rm ext}(\bm{\lambda}_Z) \odot \mathsf{T}_{\rm clone}(\bm{\lambda}_Z)
\end{eqnarray}

It is only at this stage that we may apply the instrumental broadening kernel.  The instrumental resolving power, $R$, acts as a convolution with a Gaussian line profile of width $\sigma=\frac{c}{2.355 R}$.  Real astronomical instruments usually have wavelength-dependent resolving power, which complicates the implementation for high-grasp spectra.  The extent to which this effect matters will depend on the science application.  For now, \texttt{blase} simply assumes a fixed resolving power.

Notice that the order of Steps 3 and 4 in Figure \ref{blase_flowchart} cannot be swapped.  Mathematically speaking, elementwise multiplication and convolution do not commute.  While the distinction may seem negligible, it matters at the level of precisions sought in EPRV applications (Suvrath Mahadevan, Arpita Roy, Sharon Xuesong Wang \emph{priv. comm.}).  Water vapor lines in our own atmosphere can ``beat'' with water vapor in the spectrum of, say, an M-dwarf atmosphere.  The systematic telluric mis-cancelation would imbue a Moir\'e pattern of residuals that is most acute for sources with sharp lines, namely low projected rotational broadening ($v\sin{i}\sim\frac{c}{2R}$).  The approach in \emph{blas\'e} may therefore unlock a level of telluric calibration that has evaded previous efforts.

As noted, typical data-pixel sampling $\bm{\lambda}_D$ is much coarser than the model pixel sampling $\bm{\lambda}_Z$.  We, therefore, resample the model to the data spectrum in the following way.  We evaluate the joint model $\mathsf{M}_{\rm joint}$ at all of the super-resolution wavelength coordinates and then compute the mean value of those pixels within the bounds of each coarse data pixel.  The resampling procedure is autodiff-aware: the same clusters of high-resolution coordinates map to the same data pixel coordinates, no matter what the $RV$ is.  The $RV$ only dictates what flux values are realized within those pixel bounds.

The final forward model for \emph{blas\'e} is designated simply as $\mathsf{M}$ without subscripts to emphasize that we have achieved the desired goal of a plausible end-to-end physics-informed yet highly flexible forward model for each datum in the 1D observed spectrum:

\begin{eqnarray}
    \mathsf{M}(\bm{\lambda}_D) = \resample{\Big[\mathsf{M}_{\rm joint}(\bm{\lambda}_Z) * g(R) \Big]} \label{eqn_final_model}
\end{eqnarray}

\noindent where $g$ is the Gaussian instrumental convolution kernel and the \texttt{resample[]} operation indicates the average of model pixels that fall within each data pixel's red and blue boundaries.

\subsection{Regularization}

Equation \ref{eqn_final_model} has $\sim21,000$ tunable parameters from the star, $\sim9,000$ tunable parameters from the Earth's atmosphere, plus $v\sin{i}$ and $RV$.  That adds up to about $30,002$ model parameters.  The resolving power may also be treated as tunable if it is not known or varies slightly with \emph{e.g.} seeing, slit-or-fiber illumination, or instrumental configuration: $30,003$.  The pseudocontinuum polynomial term $\mathsf{P}$ can also be tuned to refine the continuum placement initialized at the pre-processing stage, yielding an additional 5-15 parameters, depending on the choice of polynomial order.

It may appear desirable to simply optimize all of these parameters in a \emph{laissez-faire} manner, allowing them all to take on whatever value the data dictates.  Such a stratagem would overfit the data, resulting in unphysically perverse lineshapes that do not reflect the air of reality we aspire to impose on our synthetic spectral models.  Lines would haphazardly fit noise spikes, and conspire together to warp spectral shapes in unexpected ways.  This overfit model may suit some rare purposes.  But most of the time, we prefer to strike a better balance in the bias-variance tradeoff.

We apply some amount of \emph{regularization}, a restriction on the allowed values the model parameters can take on.  Fortunately, we have a firm theoretical basis to justify this regularization.  We believe our precomputed synthetic spectral models are \emph{quasi-statically correct}: the predicted spectra resemble the unobserved ``True spectrum'' with lines in the correct place, but just with the wrong area under the curve.  This statement may stem from the fact that it is easier to predict the mere existence of some energy transition of atoms and molecules than it is to predict their transition rates, abundances, temperature and pressure effects, and all the other line strength effects that flow down to how much light a line ultimately absorbs in a stellar atmosphere.

The degree of regularization will control the extent of overfitting or underfitting. The most extreme regularization---the antithesis of the \emph{laissez-faire} scenario---would yield a model too rigid to respond to the data at all, yielding a model entirely unchanged from the cloned PHOENIX and \texttt{TelFit} models, the extreme end of underfitting.  So regularization constitutes the only hyperparameters worthy of tuning in \emph{blas\'e}.  The choice of how to set the regularization is problem-specific.  We default to the following choice. We fix all line parameters except for amplitude, which receives an L1 loss-- namely, we penalize the absolute value of departures from an amplitude's starting place:

\begin{eqnarray}
    \mathcal{L}_{reg} &\equiv& \sum_{j=1}^{N_\mathrm{lines}} \Big|\frac{\ln{a_j} - \ln{\hat{a}_j}}{\Lambda}\Big|
\end{eqnarray}

\noindent where the hat notation demarcates the amplitudes obtained from cloning, \emph{i.e.} the ``initial, theory-inspired amplitude''.  We assign $\Lambda\sim5$.  The total, overall loss then becomes:

\begin{eqnarray}
    \mathcal{L}_{tot} = \mathcal{L}_{MSE} + \mathcal{L}_{reg}
\end{eqnarray}

This weak degree of regularization has the effect of permitting refinement of only the most conspicuous data-model mismatches; the weakest lines do not bother to move from their initial state because doing so would penalize the regularization without enough reduction of the overall loss.  The extrinsic $v\sin{i}$ and $RV$ have no regularization, but in practice, they need to be initialized close to their plausible values.

\section{Results: Comparison to data}\label{secResults}
\subsection{WASP 69 with HPF}

The planet-host star WASP 69 makes a great benchmark because it has a low $v\sin{i}=2.2\pm0.4~\text{km/s}$ \citep{2017A&A...608A.135C}, making its lines sharp and easy to perceive.  The K5 dwarf has an effective temperature of about 4700 K, $\log{g}=4.535\pm0.023$, and slightly super-solar metallicity \citep{2014MNRAS.445.1114A}.

\begin{figure}[hbt!]
    \centering
    \includegraphics[width=0.98\columnwidth]{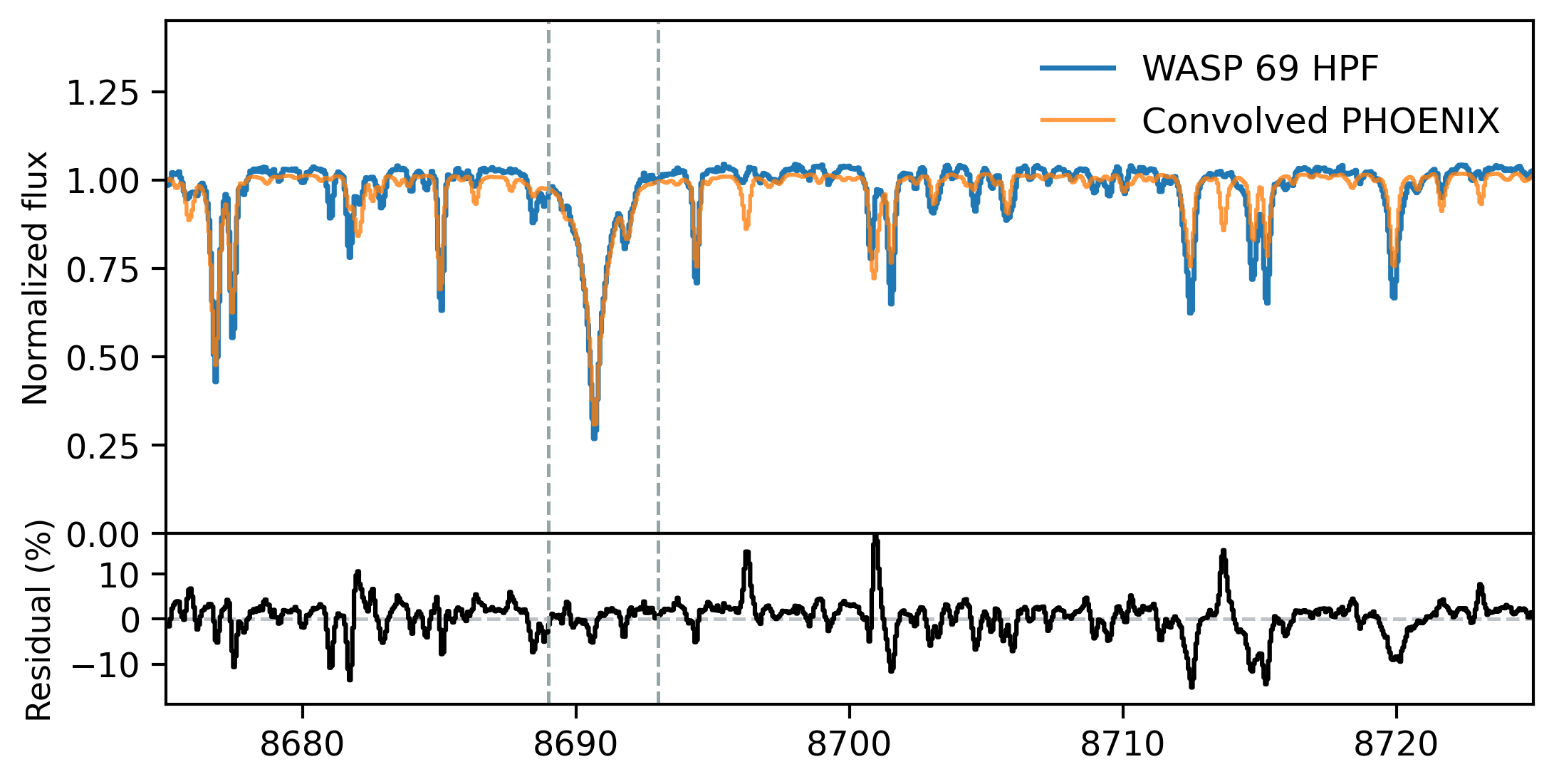}
    \caption{WASP 69 observed with HPF, compared to a $T_{\mathrm{eff}}=4700\;$K,  $\log{g}=4.5$ solar metallicity PHOENIX model warped to $v\sin{i}=2.2$~km/s, $RV=-9.6$~km/s, and HPF resolving power.}
    \label{fig_WASP69_demo}
\end{figure}

Figure \ref{fig_WASP69_demo} shows a portion of an HPF spectrum of WASP 69 centered on a region devoid of telluric lines.  This figure highlights a baseline case with no line-by-line fitting, simply conventional template matching: comparing HPF data of WASP 69 to the closest PHOENIX template, convolved and resampled to the HPF resolution and sampling.  Conspicuous residuals of $\pm10\%$ appear throughout the spectrum, with lines in the correct place, but with the amplitudes systematically biased.  These line residuals arise from bona-fide imperfections in the PHOENIX spectrum, with a minor contribution from the coarse sampling of the PHOENIX grid.

\begin{figure}[hbt!]
    \centering
    \includegraphics[width=0.98\columnwidth]{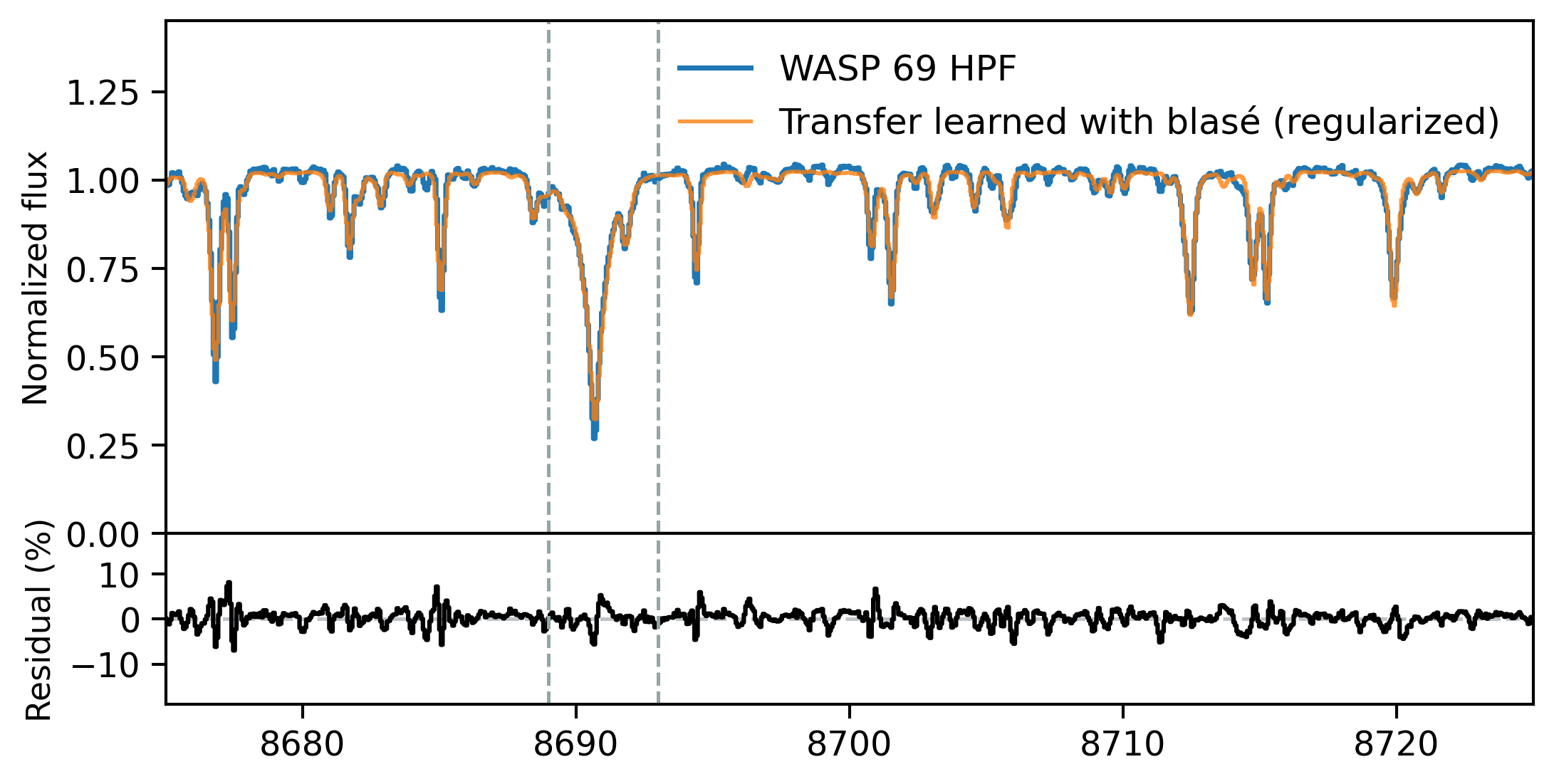}
    \caption{Semi-empirical model of WASP 69 transfer-learned with \texttt{blas\'e}, employing a regularization prior on the learned amplitudes.}
    \label{fig_WASP69_transferred}
\end{figure}

Figure \ref{fig_WASP69_transferred} shows the same data spectrum compared to a pixel-level model trained using the \emph{blas\'e} technique described in Section \ref{transferLearn}.  The model fit appears much better, with typical residuals approaching the photon noise of the data themselves.  The model is not perfect however, especially around line cores.  These residuals stem from a combination of causes.  First, our finite regularization restricts line amplitudes and widths from straying too far from their values.  This computational tug-of-war makes line cores land just short of the values they otherwise would have obtained in the absence of regularization.  Second, the spectral resolution kernel may get biased from the domineering model-misspecification of broad lines, setting up a slightly sub-par performance for all the other lines.

\begin{figure}[hbt!]
    \centering
    \includegraphics[width=0.98\columnwidth]{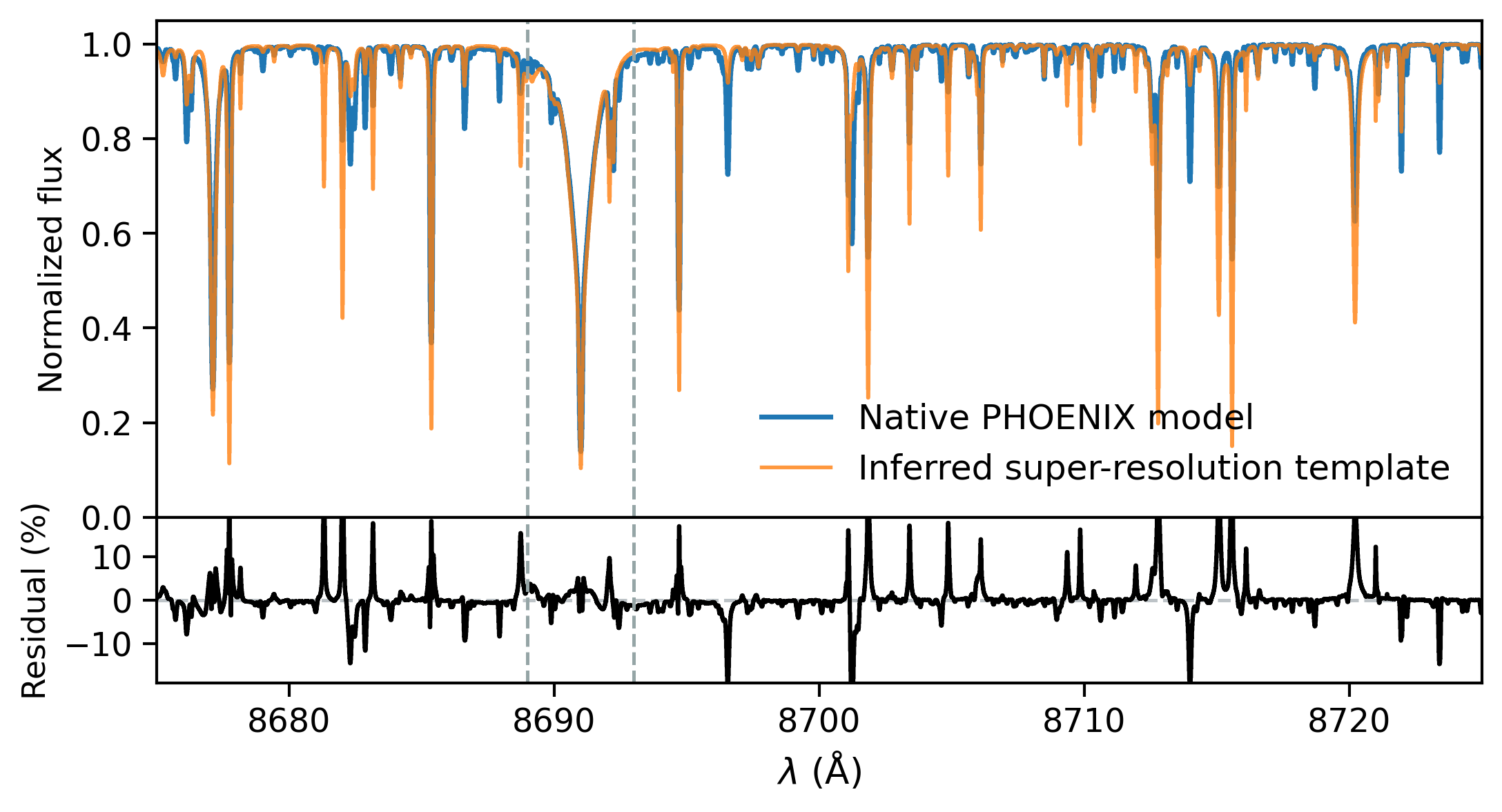}
    \caption{The native-resolution semi-empirical model transfer-learned from WASP 69 HPF data. The revised model can be viewed as a super-resolution deconvolution of the HPF spectrum.}
    \label{fig_WASP69_regularized}
\end{figure}

The inferred, semi-empirical high-resolution model for WASP 69 is therefore the transfer-learned model $\mathsf{\hat{S}}_\mathrm{clone}$, unadorned with the extrinsic and instrumental properties.  This semi-empirical model is shown in Figure \ref{fig_WASP69_regularized}.  The departures from the PHOENIX model appear dramatic at this native resolution.  The transfer-learned semi-empirical template exhibits both deeper and shallower lines than the native PHOENIX model.

\begin{figure*}[hbt!]
    \centering
    \includegraphics[width=0.95\textwidth]{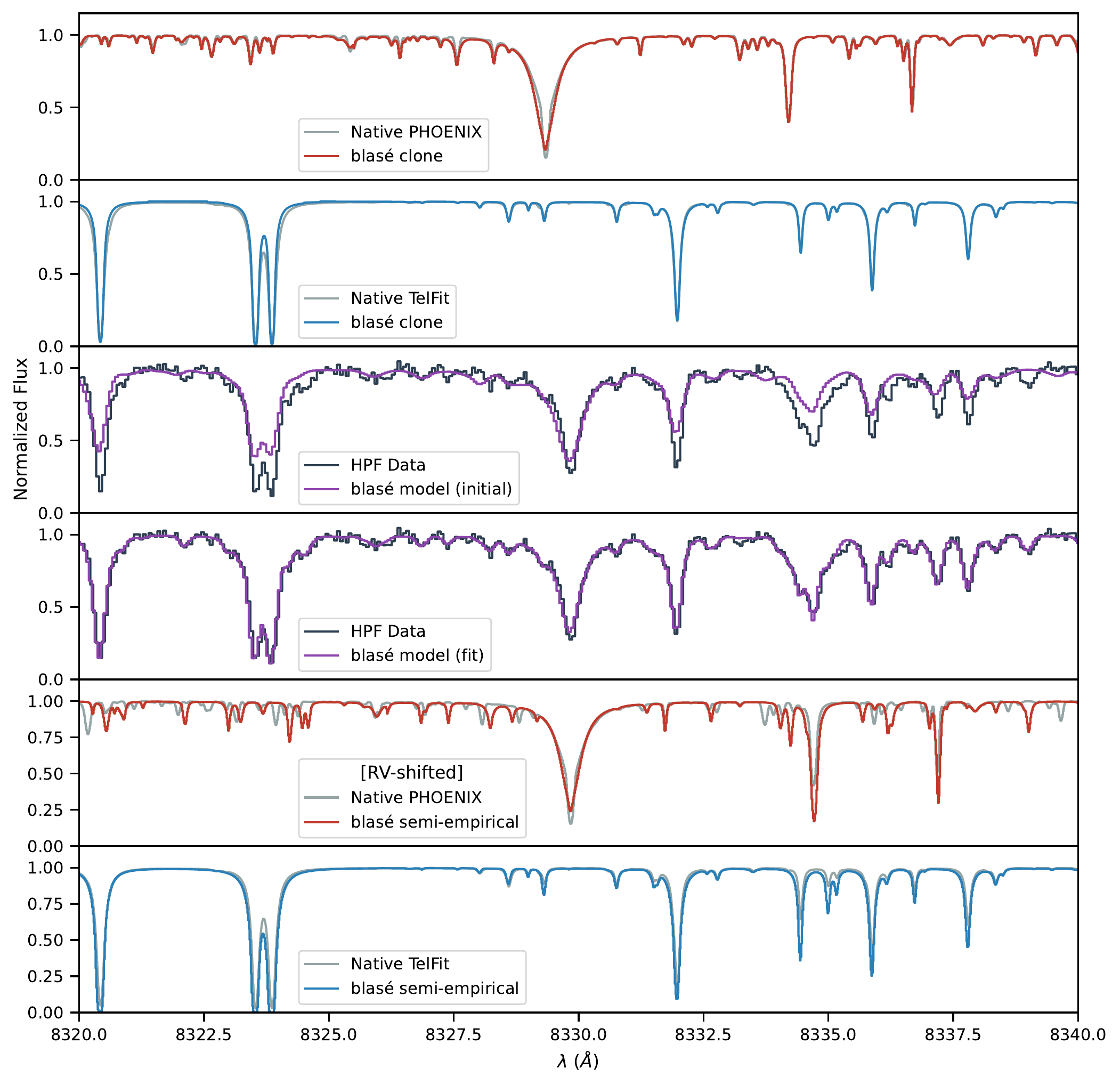}
    \caption{End-to-end training on a portion of HPF spectrum of WASP-69. The top two panels show the cloning from Step 1 of the visual guide.  The middle two panels depict the before-and-after of the instrument-convolved and resampled joint model illustrated in Step 4 of the visual guide.  The final two panels show the underlying semi-empirical models learned in the process.}
    \label{fig_multi_panel_WASP69}
\end{figure*}

The low $v\sin{i}$ of WASP 69 means that the line profile broadening arises principally from the finite instrumental resolution.  The interplay of telluric lines and stellar lines within an instrumental resolution element is exactly one of the challenges  \emph{blas\'e} was designed to solve, as discussed in Section \ref{sectionTelluric}.  In Figure \ref{fig_multi_panel_WASP69} we show a multi-panel dissection of the HPF spectrum in a wavelength region in which conspicuous stellar and telluric lines coexist.  The  \texttt{blase} end-to-end model exhibits excellent agreement with the data.  The trustworthiness of the line properties inferred in locations where stellar and telluric lines exactly overlap remains an open research question: a tie-breaker is needed to distinguish these overlapping line inferences.  We discuss conceivable tie-breakers later in the paper.

\subsection{IGRINS spectrum of a T6 Ultracool Dwarf}

The T6 ultracool dwarf 2MASS J08173001$-$6155158 was recently observed with IGRINS, revealing a rich spectroscopic atlas of molecules in its $T_\mathrm{eff} = 1060 \pm 50$ cloud-free atmosphere \citep{2022MNRAS.tmp.1421T}.  The sea of molecular lines is so rich as to blur entirely the notion of isolated lines.  Many lines should instead be considered pseudo lines.  The boundaries of lines for 2MASS J08173001$-$6155158 become even more amorphous in the presence of its moderately high $v\sin{i} = 22.5\pm 0.5 ~ \text{km/s}$ rotational broadening \citep{2022MNRAS.tmp.1421T}.  This T6 IGRINS spectrum, therefore, represents an extreme test-case, in which the underlying notions of \emph{blas\'e} go outside the comparatively safe assumptions in stellar spectra.  The application of  \emph{blas\'e} to 2MASS J08173001$-$6155158 can be viewed as conducting a deconvolution step simultaneously paired with line-by-line inference.

\begin{figure*}[hbt!]
    \centering
    \includegraphics[width=0.98\textwidth]{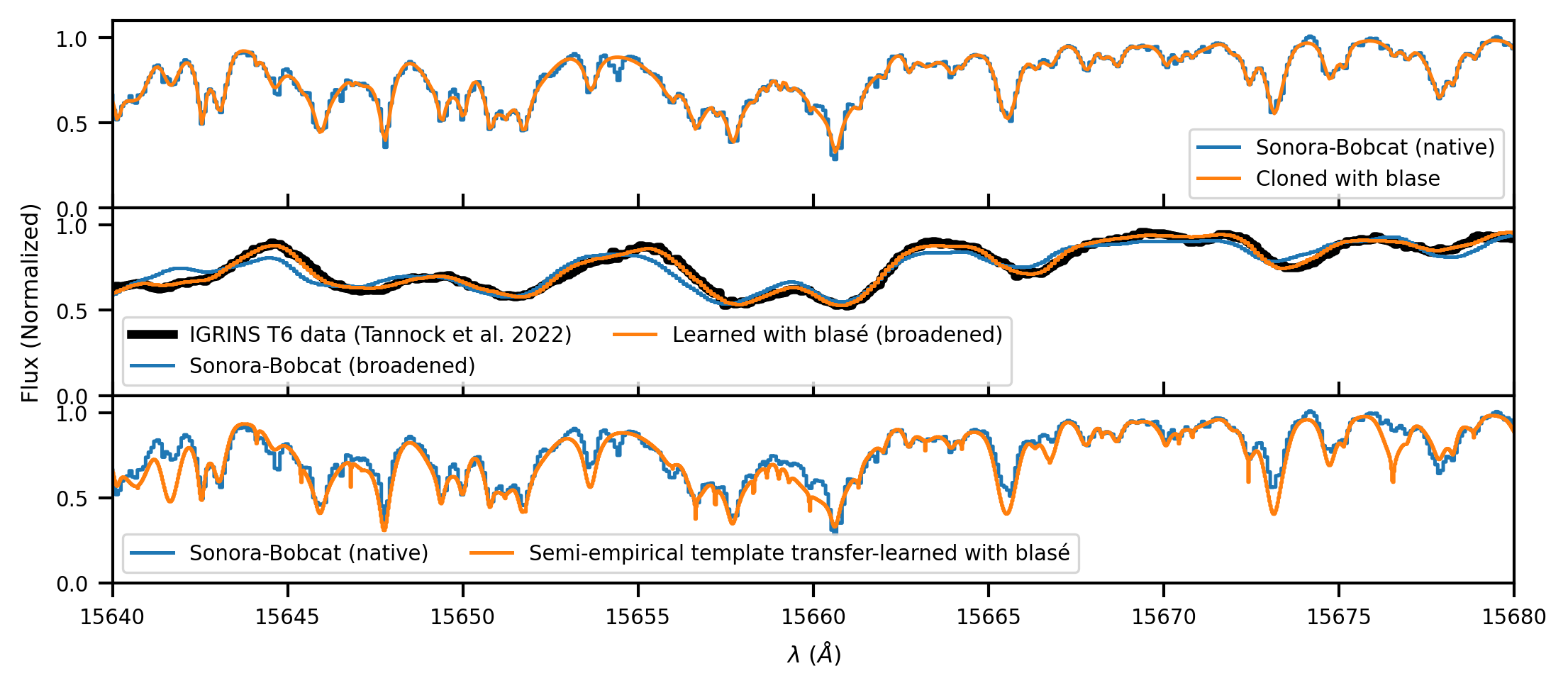}
    \caption{Line-by-line dissection of the 2MASS J08173001$-$6155158 IGRINS spectrum.  The \emph{top} panel compares the native Sonora-Bobcat spectrum to the blas\'e clone.  We match the RV shift of the observed spectrum (thick black line) and convolve the initial and learned Sonora template to the large $v\sin{i}$.  The bottom panel shows the locations and extents of line mismatches at the native resolution of Sonora.}
    \label{fig_IGRINS_sonora_demo}
\end{figure*}

We initialize \texttt{blase} with the nearest Sonora template with $T_\mathrm{eff} = 1100~\text{K}$, $\log{g}=5.0$, and solar metallicity \citep{2022MNRAS.tmp.1421T}.  Figure \ref{fig_IGRINS_sonora_demo} shows the clone of this native spectrum near the peak of the $H-$band. The cloning performance appears adequate to capture most of the molecular pseudo-lines, with a few notable shortcomings. The high signal-to-noise ratio of this IGRINS spectrum lays bare the departures from the broadened Sonora-Bobcat model, albeit with surprisingly close agreement given the challenge of substellar atmosphere modeling.  The line-by-line departures from Sonora exceed any minor flaws from the cloning process, indicating genuine opacity differences between the data and Sonora template.  Our technique automatically detects and reports the underlying structure of data-model mismatches, a key milestone for the assembly of refined opacity tables in this cool dwarf temperature regime.

\section{Discussion}\label{secDiscuss}

\subsection{How to interpret line-by-line properties}

Up to this point, we have presented the mechanics of  \emph{blas\'e}, and have shown that it applies to a wide range of applications including stars, substars, and Earth's atmosphere. The interpretation of the approach and its line-by-line outputs may carry different meaning to different practitioners.  Here we caricature some of these perspectives:

\emph{``Reverse-engineering the line lists''}-- Equation \ref{equation1} (and its exponential twin Equation \ref{equationOpacitySum}) could have been populated with initial line-by-line properties from atomic and molecular line lists, instead of the coarse detection-threshold approach we presented in Section \ref{subsecInit}.  So our cloning procedure can be viewed as reverse-engineering the quantum mechanical properties housed in line lists such as HITRAN, HITEMP, EXOMOL, VALD, or countless other primary source documents.  This interpretation is only partly accurate.  The line lists generally store the temperature and pressure scaling terms that get included in 1-D thermal structure calculations with distinct Voigt coefficient terms computed for each layer in the stellar or substellar or Earth atmosphere.  The cloned properties, therefore, represent the flux-weighted average line property, having undergone an integral through the radiative transfer in the atmosphere.

Our deliberate choice not to initialize from the atomic and molecular lines means we do not have to compute an expensive multi-layer atmosphere at each inference step.  The precomputed models have already taken care of that step, and so our cloning procedure encodes all the quantum mechanics and stellar atmospheres knowledge written down at the time of the computing of PHOENIX, Sonora, or LBLRTM.

\emph{``Equivalent Width machine''}-- The area under the curve of each $j^{th}$  \emph{blas\'e} line represents its (neighbor-weighted) Equivalent Width (EW).  That means that  \emph{blas\'e} can automatically and quickly distill a massive high-bandwidth spectrum into a catalog of line positions and EWs.  Even more, the catalog would have some resilience to rotational broadening, line blending, and other friction points that hamper such an industrial-scale approach with existing tooling.  Current applications of Equivalent Widths rely on either isolated lines or---for multi-epoch applications---the assumption of non-variable neighboring lines.   \emph{Blas\'e} does not have these limitations.  The interpretations of these EWs may have to be treated with care if intercompared with EWs obtained conventionally.  This approach may suit stellar abundance applications, such as Galactic archaeology and chemically peculiar stars \citep[\emph{e.g. HgMn stars,}][]{2020MNRAS.496..832C}.

\emph{``Fancy deconvolution''}-- The  \emph{blas\'e} approach can be viewed as a fancy deconvolution procedure, in which the convolution kernels get inferred alongside the imperfections of the model template.  This perspective may suit practitioners who wish to obtain a robust rotational broadening estimate in the presence of imperfect templates, and vice versa: robust templates in the presence of imperfect rotational broadening.

\emph{``Fancy interpolator''}--  A linear interpolator has to store the slope, offsets, and start and end points of all the line segments from pixel-to-pixel, merely to recreate a spectrum losslessly at some arbitrary coordinate or coordinates.  \emph{Blas\'e} lossily recreates a spectrum, while having to store less information.  In this way, it is fair to think of \emph{Blas\'e} as an interpolator that happens to have an intuitive interpretation.

\emph{``Surrogate modeling / Simulation Intelligence''}-- Our slightly off-label use of machine learning straddles a boundary of machine learning and conventional physics-based models.  The rapidly evolving field of machine learning may therefore classify the approach here using a few related terms.  The phrases surrogate modeling, emulation, discrepancy modeling, transfer learning, simulation intelligence, and physics-informed machine learning could describe some aspects of \emph{blas\'e}, depending on the ultimate application by the end-user.  The surrogate modeling approach has been applied with mixed success for M-dwarfs \citep{2020A&A...642A..22P} and brown dwarfs \citep{2020PASP..132d4502J}.  The realized speedups in quantum chemistry can approach 300,000$\times$ \citep{pmlr-v70-gilmer17a}. The key idea is that information from expensive physics-based simulations can be distilled by machine learning and then applied cheaply to new examples.  The performance of these surrogates should improve when more scientific domain knowledge is included at training time \citep{2018ApJ...869L...7A}.

\emph{``Narrowing the Synthetic Gap'}-- Our approach to pretraining on synthetic spectra and comparing to data resembles the technique of \citet{2020A&A...642A..22P}. There, a convolutional neural network is pretrained on PHOENIX spectra, and then applied to CARMENES spectra of M dwarfs.  \citet{2020A&A...642A..22P} designate the difference in appearance between the synthetic spectra and observed spectra as the ``synthetic gap'', showing that this gap leads to inaccurate metallicity inferences.  \emph{Blas\'e} shows a path towards shrinking the synthetic gap, with the prospect of re-training convolutional neural networks with semi-empirical spectra produced via the method described here, assuming high-fidelity metallicity labels can be obtained through co-natal binary studies.

\emph{``A good enough substrate''}-- The imperfections in pre-computed stellar and substellar models have hampered their use in high-fidelity applications, where model flaws overwhelm the signal-to-noise in the data.  The adaptability of  \emph{blas\'e} can be viewed as increasing the number of applications that could conceivably benefit from the rich-albeit-imperfect information encoded in these model grid spectra.  In the next Section, we illustrate how these adaptable models can serve as a substrate for applications in which the imperfections in precomputed models would have otherwise precluded their use.

\subsection{PyTorch and the autodiff ecosystem}

One key limitation stems from the need for autodiff to ``know about'' or sense each step of the end-to-end computation.  Practically speaking, this requirement means \texttt{blase} had to be written in a single differentiable (\emph{i.e.} autodiff-aware) programming framework.  This requirement is so strong as to forbid the intermingling of frameworks and languages.  The choice of which machine learning framework to adopt must therefore be made up-front, with some risk of \emph{vendor lock-in} once that choice has been set.  We considered three such differentiable frameworks: \texttt{TensorFlow}, \texttt{JAX}, and \texttt{PyTorch}.  We chose \texttt{PyTorch} because of its ease of use and its ability to structure differentiable models in an object-oriented style familiar within the \texttt{Python} community, with robust optimization support.  PyTorch had already been applied to several astronomy projects by this time, including \texttt{Exonet} \citep{2018ApJ...869L...7A}, \texttt{PySR} \citep{pysr}, and \texttt{MPoL} \citep{mpol}.  Ultimately, the choice of \texttt{PyTorch} made it easy to employ Sparse Matrices.

This vendor lock-in gives rise to a frustrating incompatibility among exoplanet-relevant autodiff applications.  For example, no differentiable piece of code in \texttt{blase} can call an \texttt{exoplanet} orbital radial velocity, a \texttt{starry} surface map, an \texttt{astropy} spectral coordinate transformation, a \texttt{wobble} spectral residual model, or an \texttt{ExoJAX} Voigt profile.  All of those tools use either \texttt{Theano}, \texttt{TensorFlow}, \texttt{Numpy}, or \texttt{JAX}, which are incompatible with \texttt{PyTorch}.  We use \texttt{SciPy}, but only for preprocessing, before the differentiable model has been entered.  Some heuristics exist for interchanging code among the frameworks, but they pose some implementation complexity and/or some performance hit.

It may appear attractive to coordinate the relatively small autodiff-for-astrophysics community to adopt a single autodiff-able machine learning framework to facilitate compatibility.  In practice, such coordination may be ill-fated, since the rapid pace of machine learning development has not yet settled on a single best way forward.  Each framework offers some strengths and demerits, and exploring those tradeoffs will continue to yield new experimental approaches.  For example, \texttt{JAX} may offer speedups attributable to Just In Time (JIT) compilation, at the expense of a more rigid functional programming style.  Access to scalable Gaussian Process implementations may be another deciding factor \citep{2022arXiv220908940A}.

\section{Unlocking new science}\label{secFutureWork}

The scaffolding of \texttt{blase} is designed in a way to promote extensibility, so many scientific questions can be written down in the flexible language presented in this paper.  In some cases, the flexibility, precision, speed, and ease of use may unlock new approaches to long-standing astrophysical questions and practical challenges. Here we enumerate some planned or conceivable extensions.  We break up the themes into two categories: scientific extensions (this Section) and technical innovations (Section \ref{technicalInnovations}).

\subsection{Extreme Precision Radial Velocity} \label{eprv}

Already \texttt{blase} is equipped to fit every single line with its own systemic $RV$, by tuning the line center position $\lambda_\mathrm{c}$ at each training epoch.  There exists both empirical evidence and some theoretical motivation that $RV$ jitter varies from line-to-line \citep{2018A&A...620A..47D}.  Importantly, the extent of this line-by-line RV jitter could be predicted in part by the depth of line formation \citep{2021A&A...653A..43C, 2022A&A...664A..34A}.  A future extreme-precision-RV (EPRV) version of \texttt{blase} could leverage this information.  As an example, the depth-of-formation for all (or a subset) of lines could be obtained and associated with each spectral line.  A regularization scaling term could be introduced to allow the line positions to vary, but only in proportion to their depth of formation.

\subsection{Line-by-line fundamental parameter estimates}
The conversion of spectral line strengths to spectral type \citep{1901AnHar..28..129C}---and by extension line strengths to $T_\mathrm{eff}$ \citep{1925PhDT.........1P}---has occupied a large chunk of stellar spectroscopy in the last century.  The measurement of fundamental stellar properties remains one of the principal applications of stellar spectroscopy.   \emph{Blas\'e} as it currently stands does not output fundamental properties, and in fact it takes them \emph{for granted}: you must specify \emph{e.g.} the $T_\mathrm{eff}$, $\log{g}$, [Fe/H], and [$\alpha$/Fe] of your PHOENIX template as Step Zero before that template gets cloned and warped to match data.  So obtaining a $T_\mathrm{eff}$ estimate from  \emph{blas\'e} is impossible, at least as discussed so far.

The clearest way forward would be to calibrate the blas\'e-derived line properties.  Simply put, we seek a set of functions, $f$, that relate the line properties to fundamental properties:
$$f_j(a_j, \sigma_j, \gamma_j) \to (T_\mathrm{eff}, \log{g}, [\mathrm{Fe}/\mathrm{H}])$$

There are $N_{\mathrm{lines}}$ such functions---one for each spectral line---because each line has its own temperature, gravity, and metallicity dependence.  For visualization simplicity purposes, we have assumed the PHOENIX grid is merely 3-dimensional ($T_\mathrm{eff}$, $\log{g}$, [Fe/H]), but the same logic applies to 4-D and higher, and it works for sparse and irregularly sampled grids too.

Fascinatingly, there are two related ways forward towards this function.  They both involve ensembles of spectra---such as a spectral atlas, spectral sequence, or library---and they both involve first obtaining the inverse function:

\begin{eqnarray}
    g_j &\equiv& f^{-1}_j\\
    g_j(T_\mathrm{eff}, \log{g}, [\mathrm{Fe}/\mathrm{H}]) &\to& (a_j, \sigma_j, \gamma_j)
\end{eqnarray}

A purely model-based approach would start by  \emph{blas\'e}-cloning every single PHOENIX spectrum in the grid dimensions-of-interest, over the wavelength range-of-interest.  The resulting product would be a line-by-line catalog of cloned spectral properties for each 3D coordinate on this grid.  We could then assemble a heatmap of how each $j^{th}$ line property changes across this heatmap, $g_j$.  In this way, we are reverse-engineering the \emph{e.g.} temperature dependence of each spectral line, as encoded by the PHOENIX atmosphere models.  The final step could involve finding the nearest neighbor of each $j^{th}$ observed spectral line to each $j^{th}$ grid heatmap point; that is, finding a way to invert the function $g_j$ to get $f_j$.  The information-weighted mean, median, or mode of these nearest neighbors could then be reported as a revised ``best-fit $T_\mathrm{eff}$'', for instance.

A semi-empirical approach could improve on this purely model-based approach.  We know that the PHOENIX line depths, widths, and shapes are imperfect, and so this ($T_\mathrm{eff}$, $\log{g}$, [Fe/H]) heatmap will have large flaws in the line-by-line properties: the contours are systematically too bright or dark.  There are many conceivable ways to quantify these heatmap flaws.   \emph{Blas\'e} provides an expedient route.  One can pull the heatmap towards the locus of points established by running  \emph{blas\'e} on benchmark stars.  The known $T_\mathrm{eff}$ of such systems would anchor the trend.  We suspect the directionality of the purely model-based heatmap must be correct, simply the slope, offset, and concavity of the trends may be wrong. With new information from each benchmark, the heatmap would get lifted like a central tentpole propping up an under-supported tent.

The construction and calibration of these ensembles of spectra represents a tremendous amount of work beyond the scope of this paper.  But its creation could yield an extremely precise, fast, interpretable, and reasonably accurate way to measure the properties of stars based solely on their high bandwidth \'echelle spectra.  The overall accuracy hinges on the accuracy of the stellar benchmark labels.  The mechanics of this approach could be adapted to fit within existing pipelines such as SAPP \citep{2022A&A...658A.147G}.

There exists a corollary from the method described here to the approach of \citet{czekala15}.  There, a 3D heatmap was created for relating the eigenweights of a PCA basis to the stellar fundamental properties, for spectrum emulation purposes.  Here, the 3D heatmap is created on a line-by-line basis and is therefore interpretable.  The PCA eigenspectra were generally uninterpretable, at least not easily.  So in the limiting case of obtaining a densely calibrated semi-empirical heatmap for all spectral lines across all grid dimensions, we will have achieved a spectral emulator that would obviate the need to clone spectra in the first place.  We would simply start with this powerful line-by-line emulator as a forward model, and go directly to the extrinsic warping and data-model comparison.

\subsection{Abundances}\label{secAbundance}

Stellar abundance work involves precisely measuring line strengths (often reported as EWs) of different chemical constituents evinced in a star or substar's photosphere.  Relating those EWs to physics can be done in a few ways.  Most easily, trends and patterns in the EWs can be assembled, and metal-rich and metal-poor clusters can be identified.  There exists a precision/accuracy tradeoff in the measurement process, usually stemming from the placement of the continuum, or assumptions about line blending.   \emph{Blas\'e} offers an immediate solution to these challenges since it starts from our best guess for how nearby lines may be shaping the continuum.   \emph{Blas\'e} appears to be a gateway to extremely fast and intervention-free industrial-scale abundances, potentially useful for large surveys like APOGEE \citep{2017AJ....154...94M}, Gaia-ESO \citep{2012Msngr.147...25G}, RAVE \citep{2006AJ....132.1645S}, and more.

\subsection{Identifying missing lines}\label{secMissingLines}
So far we have not addressed the inevitable prospect that some spectral line or lines reside in the observed data spectrum, but are absent entirely from the precomputed model.  This ``missing line'' scenario arises from our incomplete and evolving knowledge of quantum mechanics and molecular chemistry.  \texttt{ROBOSPECT} \citep{2013PASP..125.1164W} and the \texttt{Cannon} \citep{2017ApJ...836....5H} have data-driven ways to identify these lines.  The \texttt{Cannon} can go further to indicate how these previously unknown lines correlate with the stellar labels, providing some physical understanding of their cause. \emph{Blas\'e} as it currently stands has no mechanism for identifying these lines.  Here we consider that there are $N_\mathrm{missing}$ lines with two scenarios: \emph{I)} the identification, line center location, and possibly other ancillary information about the lines have become newly available,  or \emph{II)} the lines are truly anomalous and unknown.

For the first scenario, Voigt profiles could be instantiated at the newly-established line center locations $\lambda_{c,m}$ of the previously unknown line.  So Equation \ref{equation1} would become a product of the new and old physics:

\begin{eqnarray}
    \mathsf{S}_{\rm clone} = \Big( \prod_{m=1}^{N_{\mathrm{missing}}} 1-a_m \mathsf{V}_m \Big) \Big(\prod_{j=1}^{N_{\mathrm{lines}}} 1-a_j \mathsf{V}_j \Big) \label{missing}
\end{eqnarray}

\noindent Scenario \emph{II}---truly unknown lines---would not have the luxury of knowing where to initialize the line center location, at least not based on precise theory.  So instead, an educated guess for $\lambda_\mathrm{c,m}$ could be made based on the data centroid.

In either case, the initialization of $A_m, \sigma_m, \gamma_m$ would be a matter of taste, and they would therefore be difficult to regularize, at least in the manner that we have described so far.  For these reasons, an even better strategy may be to treat the data residuals not as lines, but in some other basis.  For example, it could be possible to model the \emph{blas\'e} residual spectrum with, say, \texttt{wobble}, or with a Gaussian process.  These questions remain an open research area.

\subsection{Doppler Imaging}
The fixed $v\sin{i}$ approximation breaks down for stars with large-scale surface features.  Doppler imaging attempts to infer the surface map from the extent to which observed line profiles depart from a pristine rotational broadening kernel.  This inference procedure suffers from a vast number of geometrical degeneracies, but still provides useful constraints on stellar surfaces \citep{2021arXiv211006271L}.  We emphasize a distinction between A) longitudinally symmetric surface features, and B) longitudinally asymmetric surface features.  Most radial velocity practitioners think about the latter, since longitudinally asymmetric surface features imbue changing-in-time skewness to the line profiles, causing radial velocity perturbations easily detectable in radial velocity time series.  These confound exoplanet searches.

Longitudinally symmetric surface features---on the other hand---do not change as the star rotates on its axis.  The existence of these features manifests as static-in-time \emph{kurtosis} of the spectral line.  For example, a hypothetical non-emitting (black) polar starspot exhibits a deficit in flux at the line core, resulting in less zero-velocity flux than its homogeneous counterpart.  A dark zonal band results in equal-sized bites out of the red- and blue- sides of the line.

It is easier to infer longitudinally asymmetric features than longitudinally symmetric ones, since we assume that we occasionally catch a glimpse of the spot-free limbs and their pristine line profiles.
The latter requires exact knowledge of the underlying spectral template.  Isolated, deep, well-calibrated spectral lines constitute the only practical scenario where exact knowledge can plausibly be claimed.  Isolated spectral lines may be scarce or absent for M-dwarfs and brown dwarfs where lines blend ostensibly in an inseparable way, confounding Doppler imaging.

\emph{Blas\'e} offers a new approach to Doppler imaging that may overcome these historical limitations by simultaneously fitting both the imperfections in the underlying spectrum and its line profile perturbations.  This approach is analogous to the linearized model in \citet{2021arXiv211006271L}, but with the benefit of also handling non-linear properties of the spectrum such as line widths, shapes, and locations, while also handling telluric contamination.  Hypothetically the \citet{2021arXiv211006271L} approach could be partially absorbed into \texttt{blase}, or vice versa, though such a merger may be complicated to implement.

\subsection{Starspots and magnetic fields}
One current assumption of starspot spectral decomposition is that the starspot spectrum itself resembles the stellar photosphere of a cooler star.  This assumption appears adequate for detecting starspot spectra and measuring their physical properties \citep{2017ApJ...836..200G}.  But to second order, starspots should exhibit some spectral peculiarities that make them depart from a ``normal'' stellar photosphere.  We may be probing deeper into the photosphere, and so the lines may experience higher pressure, with slightly different line widths \citep{oneal96}.  Or maybe the finite convective velocity shift can be directly seen as systematic shifts of the spectral lines (as stated in Section \ref{eprv}).

Rather than applying a mixture model of fixed PHOENIX templates, one could adopt a mixture model of pre-cloned  \emph{blas\'e} models.  Then, the imperfections in the starspot spectrum can be learned alongside the filling factor of the starspots.

Relatedly, magnetic field information could be incorporated into \emph{blas\'e} as line-by-line Land\'e $g-$factor ``labels'', when available. The use of these labels could take a few different forms. The labels could simply serve as flags, to indicate that regularization should be weakened to allow these magnetic sensitive lines to wander farther from their naive non-magnetic expectations.  Or the labels could be used to infer the extent of Zeeman broadening or Zeeman splitting by parameterizing these operations as convolutions.  The best choice for how to incorporate magnetic sensitivity into the flexibility of \emph{blas\'e} remains an open research question.

\subsection{Circumstellar disk and accretion veiling}
Circumstellar disk veiling suppresses the strengths of all spectral lines \emph{en masse}, as the stellar photosphere gets outshined by a hot disk and/or envelope.

The bulk properties of that disk/envelope could be incorporated with a simple physical model: a black body of temperature $T_\mathrm{disk}$ and solid angle $\Omega_{disk}$ \citep{2018ApJ...862...85G}.  Revealing the spectral shape of the veiling would require sufficiently high bandwidth spectra, such as Xshooter \citep{2011A&A...536A.105V} or possibly IGRINS.  The extent of veiling would be fit alongside all the other stellar spectral lines.  However, an MCMC approach may outperform  \emph{blas\'e} in accuracy, since the choice of picking an underlying spectral template is partially degenerate with the derived veiling.

Accretion veiling can be treated analogously, but with some more complications.  Emission lines can be easily incorporated as described in Section \ref{gpuConsiderations}.
These emission lines could be initialized with a line list of hydrogen lines, forbidden lines, and other conspicuous features.  The shapes of those lines can be affected by winds and other physical phenomena, yielding a variety of physically interpretable line shapes \citep{2022arXiv220802940E}.   \emph{Blas\'e} could incorporate those lineshapes as templates shared among several lines originating from similar physical environments, but scaled and shifted based on the details of each line's radiative transfer properties.

\subsection{Sky emission lines and wavelength calibration}
Night sky emission lines---arising from OH, for example---add emission line spikes on top of the astronomical target spectrum of interest.  These lines get subtracted in most pipeline packages, but in some cases, a more careful treatment may be desirable.  In HPF, a sky reference fiber points towards blank sky, needing some careful calibration before subtraction from the target fiber \citep{2022JOSS....7.4302G}.  Remote sensing applications may wish to measure such emission lines as the primary science of interest.  It would be straightforward for \emph{blas\'e} to handle emission lines, following the augmentation for emission lines in Section \ref{gpuConsiderations}.  The main benefit over conventional sky subtraction methods may be the ability to disentangle the underlying spectrum at super-resolution.  A predictive model for sky emission could be built-up over time, building data-driven corrections to theoretical models.

The well-known center wavelengths of these lines could be used for additional wavelength calibration, as is already done in IGRINS \citep{jaejoonlee16}.  An imperfect instrumental wavelength calibration could be diagnosed based on the departure of the predicted versus realized wavelength center positions, enabling a relatively fast and intervention-free correction strategy.  Hypothetically, a prescription for the wavelength calibration could be built directly into \emph{blas\'e}, with the coefficents for a Polynomial re-routing the realized wavelength center locations from their known wavelength centers.

\section{Conclusions}
We have introduced an interpretable machine learning approach to forward modeling stellar, substellar, and telluric spectroscopic data.  The line-by-line approach relies on a key enabling technology, automatic differentiation, that allows a nearly unlimited number of spectral lines to be forward modeled simultaneously.  We initialize these lines to match precomputed synthetic stellar spectra, achieving excellent performance, and lending some confidence that the approach has a capacity to capture a tremendous amount of information at once.

We demo the framework on two sources: the K5 exoplanet host star WASP 69 using a precomputed PHOENIX model, and the T6 ultracool dwarf 2MASS J08173001$-$6155158 using a precomputed Sonora-Bobcat model.  We discuss how  \emph{blas\'e} can be used to measure Equivalent Widths for thousands of lines automatically, understand line lists, measure rotation rates, generate surrogate models, and construct semi-empirical models.  This tool could readily have applications across stellar and substellar astronomy, including for PRV work, stellar compositions, Doppler imaging, and stellar activity.

\pagebreak
\newpage

\begin{acknowledgments}
    \footnotesize{
        We thank the anonymous referee for comments that improved the paper.  We thank Phill Cargile, Hajime Kawahara, and Dan Jaffe for preprint comments.

        This material is based upon work supported by the National Aeronautics and Space Administration under Grant Numbers 80NSSC21K0650 for the NNH20ZDA001N-ADAP:D.2 program,
        and 80NSSC20K0257 for the XRP program issued through the Science Mission Directorate.

        We acknowledge the National Science Foundation, which supported the work presented here under Grant No. 1910969.

        These results are based on observations obtained with the Habitable-zone Planet Finder Spectrograph on the HET. The HPF team was supported by NSF grants AST-1006676, AST-1126413, AST-1310885, AST-1517592, AST-1310875, AST-1910954, AST-1907622, AST-1909506, ATI 2009889, ATI-2009982, and the NASA Astrobiology Institute (NNA09DA76A) in the pursuit of precision radial velocities in the NIR. The HPF team was also supported by the Heising-Simons Foundation via grant 2017-0494.

        The Hobby-Eberly Telescope (HET) is a joint project of the University of Texas at Austin, the Pennsylvania State University, Ludwig-Maximilians-Universit\"at M\"unchen, and Georg-August-Universit\"at G\"ottingen. The HET is named in honor of its principal benefactors, William P. Hobby and Robert E. Eberly.

        MGS thanks Ian Czekala, Phill Cargile, Rodrigo Luger, Greg Herczeg, Will Best, Dan Foreman-Mackey, Meg Bedell, Mark Marley, Erwan Pannier, Dirk C.M. van den Bekerom, Dan Clemens, Dan Jaffe, Tom Greene, Adam Kraus, David Hogg, Greg Zeimann, Ben Montet, Christina Hedges, Brittany Miles, Arpita Roy, and Kevin Gullikson for conversations and resources that shaped his thinking on spectral calibration and telluric mitigation.
    }
\end{acknowledgments}

\facilities{HET (HPF)}

\software{ pandas \citep{mckinney10},
    matplotlib \citep{hunter07},
    astropy \citep{exoplanet:astropy13,exoplanet:astropy18},
    exoplanet \citep{2021JOSS....6.3285F},
    numpy \citep{harris2020array},
    scipy \citep{2020SciPy-NMeth},
    ipython \citep{perez07},
    starfish \citep{czekala15},
    seaborn \citep{Waskom2021},
    pytorch \citep{2019arXiv191201703P},
    muler \citep{2022JOSS....7.4302G}}

\bibliography{ms}

\clearpage

\appendix
\restartappendixnumbering

\section{Log flux scaling mode} \label{appendixLogScale}

Here we illustrate how  \emph{blas\'e} gets altered when applying the logarithmic flux pre-processing step.  First, we compute the natural log of the flux directly on the precomputed synthetic spectrum in its absolute flux scaling and native pixel sampling:

\begin{eqnarray}
    \ln{\mathsf{S}} = \ln{\mathsf{S}_{\rm abs}} - \ln{\mathsf{B}} - \mathsf{P}
    \label{eqnlogFlat}
\end{eqnarray}

We simply ``rebrand'' $\mathsf{P}$ as residing in logarithmic flux units, and disregard it since it is largely a nuisance parameter anyways.  We then treat the  \emph{blas\'e} clone model as a sum of opacities, retaining the Voigt profile:

\begin{eqnarray}
    \ln{\mathsf{S}_{\rm clone}} = -\sum_{j=1}^{N_{\mathrm{lines}}} a_j \mathsf{V}_j \label{equationOpacitySum}
\end{eqnarray}

Here, the $a_j$'s have also been slightly rebranded from their meaning in Equation \ref{equation1}.  We still want to enforce only absorption lines---and not spurious emission lines---so we use the sample trick of sampling the $a_j$'s in log and then exponentiating them to get guaranteed positive values.  Note that Equations \ref{equation1} and \ref{equationOpacitySum} carry modified meanings for the Voigt profile.  Specifically, Equation \ref{equation1} can be viewed as the Taylor Series expansion for \ref{equationOpacitySum} in the limit of small opacities:

\begin{eqnarray}
    e^{-a_j \mathsf{V}_j} \approx (1-a_j\mathsf{V}_j) \label{eqnTaylor}
\end{eqnarray}

Both equations are approximate. A real stellar atmosphere's lineshape arises from a sum of disparate Voigt profiles weighted along a nonuniform column of gas, whereas here we have assumed the column of gas is approximated as a single uniform isothermal backlit layer.  A sum of unlike-Voigt profiles is not exactly equal to any single Voigt profile.  Theoreticians may resonate with this more ``first principles'' representation, while data practitioners may find Equation \ref{equation1} more natural, so to some extent the choice is a matter of taste.

The sparse matrix gets rebranded as filled with opacity values, instead of log-fluxes, but operationally remains the same. All subsequent steps operate on the summed-and-exponentiated opacities, behaving identically to their linear counterparts.  For example, we exponentiate before computing the residuals and data-model comparison, $\mathsf{R} = e^{\ln{\mathsf{S}}} - e^{\ln{\mathsf{S}_{\rm clone}}}$.

\begin{deluxetable}{cp{10cm}}
    \tabletypesize{\scriptsize}
    \tablecaption{Notation used in this paper\label{table2}}
    \tablehead{
        \colhead{Symbol} & \colhead{Meaning}
    }
    \startdata
    \hline
    \multicolumn{2}{c}{Spectra}\\
    \hline
    $\bm{\lambda}_S$ & Native wavelength coordinates of the precomputed stellar spectrum\\
    $\bm{\lambda}_T$ & Native wavelength coordinates of the telluric spectrum\\
    $\bm{\lambda}_D$ & Native wavelength coordinates of the data spectrum\\
    $\mathsf{S}_{\rm abs}$ & Flux values of the precomputed synthetic stellar spectral model $\bm{\lambda}_S$\\
    $\mathsf{B}$ & Blackbody of temperature $T_{\mathrm{eff}}$ to coarsely normalize $\mathsf{S}_{\rm native}$\\
    $\mathsf{P}$ & Smooth polynomial to refine continuum-normalization\\
    $\mathsf{S}$ & Continuum normalized augmentation of $\mathsf{S}_{\rm abs}$\\
    $\mathsf{T}$ & Transmission values of the precomputed synthetic telluric model \\
    $\mathsf{D}$ & The observed data spectrum flux values\\
    $\bm{\epsilon}$ & The estimated uncertainties in the data spectrum\\
    $\mathsf{S}_{\rm clone}$ & Evaluable and tunable cloned flux model of $\mathsf{S}$\\
    $\mathsf{T}_{\rm clone}$ & Evaluable and tunable cloned transmission model of $\mathsf{T}$\\
    $\mathsf{S}_{\rm ext}$ & An augmentation of $\mathsf{S}_{\rm clone}$ with $v\sin{i}$ convolution and $RV$ translation\\
    $\mathsf{M}_{\rm joint}$ & The joint stellar and telluric model: $\mathsf{S}_{\rm ext} \odot \mathsf{T}_{\rm clone}(\bm{\lambda}_S)$  \\
    $\mathsf{M}$ & Joint model convolved
    with instrumental kernel and resampled to $\bm{\lambda}_D$\\
    $\mathsf{R}$ & The residual spectrum between a pair of inputs, \emph{e.g.} $\mathsf{D} - \mathsf{M}$\\
    $\bm{v}$ & The spectral coordinate axis $\bm{\lambda}$ expressed as a velocity difference\\
    \hline
    \multicolumn{2}{c}{Line properties}\\
    \hline
    $\lambda_{\mathrm{c},j}$ & Line center position of the $j^{th}$ spectral line\\
    $a_j$ & Gaussian line profile amplitude of the $j^{th}$ spectral line \\
    $\sigma_j$ & Gaussian line profile scale of the $j^{th}$ spectral line\\
    $\gamma_j$ & Lorentzian line profile half width of the $j^{th}$ spectral line\\
    $\mathsf{V}_j$ & The Voigt profile of the $j^{th}$ spectral line \\
    $\bar{\bm{F}}$ & The dense $(N_{\rm lines} \times N_{x})$ matrix of all line fluxes stacked vertically \\
    $\hat{\bm{F}}$ & The sparse $(N_{\rm lines} \times N_{\rm sparse})$ matrix of all line fluxes stacked vertically \\
    $\zeta$ & The rotational broadening convolution kernel\\
    $g$ & The instrumental broadening convolution kernel, typically a Gaussian\\
    \hline
    \multicolumn{2}{c}{Scalars}\\
    \hline
    $N_{\rm lines}$ & Number of spectral lines \\
    $N_{x}$ & Number of pixel coordinates in the precomputed spectrum $\bm{\lambda}_x$\\
    $N_{\rm sparse}$ & Number of non-zero pixels computed in the sparse implementation\\
    $\pm \Delta \lambda_{\mathrm{buffer}}$ & Buffer exceeding the red and blue limits of the data spectrum\\
    $P_{\rm rom}$ & The prominence threshold of spectral lines to include in cloning \\
    $v\sin{i}$ & Rotational broadening for stellar inclination $i$ and equatorial velocity $v$\\
    $RV$ & Radial velocity of the star\\
    $R$ & Spectrograph resolving power $\lambda/\delta\lambda$\\
    $\mathcal{L}$ & The loss scalar, usually the sum of the squares of the residuals\\
    \hline
    \multicolumn{2}{c}{Operators}\\
    \hline
    $\resample \big[ \mathsf{F(\bm{\lambda}_x)} \big]$ & The resample operator, takes in a flux spectrum $\mathsf{F}$ evaluated at $\bm{\lambda}_x$ coordinates and returns the mean flux within the pixel boundaries of coordinate $\bm{\lambda}_z$\\
    $*$& The convolution operator\\
    $\odot$& \emph{Hadamard product}, an elementwise product of two same-length vectors\\
    \enddata
\end{deluxetable}

\section{Comparison to existing spectroscopy frameworks}
Several astronomical spectral frameworks share similar aims as  \emph{blas\'e}.  These existing frameworks will have enduring value for the wide range of problems in the field of stellar spectroscopy.  Here we scrutinize the differences among some of these approaches to clarify how this work fits in.

The \texttt{specmatch} synthetic template matching tool produces noise-free nearest neighbor templates given an input spectrum \citep{2015PhDT........82P}.  Several practical barriers limit the accuracy of using precomputed synthetic spectral models alone. First and foremost, real stars are usually more complicated than our simplified models of them. Real spectra often vary over more dimensions that our models do.  Conspicuous examples of these hidden variables can be found in protostars: starspots, accretion veiling, dust extinction, and magnetic Zeeman splitting. Jointly modeling all of these phenomena alongside the intrinsic stellar photosphere is challenging.

The empirical version, \texttt{specmatch-emp} \citep{2017ApJ...836...77Y} matched spectra better than the synthetic templates, but is still too rigid for some applications and requires the assembly of hundreds of standardized high signal-to-noise-ratio templates, ideally with low intrinsic rotational broadening.  Such a large number of high-quality templates with high resolving power and low $v\sin{i}$ has not yet been established in the near-infrared.

The \texttt{wobble} framework \citep{2019AJ....158..164B} modernized the construction of high-SNR templates to account for temporally variable telluric lines. The tool requires dozens of high-SNR spectra acquired at a range of Barycentric Earth Radial Velocities (BERVs).  The final telluric-free combined spectrum would still have to be compared to models for absolute calibration or can be used out-of-the-box for precision relative RVs.  The \texttt{wobble} framework also pioneered the off-label application of automatic differentiation frameworks---in this case \texttt{TensorFlow}---towards their physically-motivated use in stellar spectra.  \texttt{blase} can be viewed as an evaluable and interpretable super-resolution version of \texttt{wobble}, that accepts more bias in the bias-variance tradeoff.

The \texttt{starfish} framework \citep{czekala15} provides a robust likelihood function for data-model comparisons and retires many of the problems in this domain.  \texttt{starfish} pioneered the use of whole-spectrum fitting with resilience to model imperfections by addressing the problem of what to do when the underlying atomic and molecular data was wrong or approximate or missing.  It has been extended to inferring starspot physical properties \citep{2017ApJ...836..200G}, measuring veiling in Class 0 protostars \citep{2018ApJ...862...85G}, and quantifying imperfections in brown dwarf models \citep{2021ApJ...921...95Z}.  The Spectral Inference Crank \citep[\texttt{sick},][]{2016ApJS..223....8C} shares similar aims as \texttt{starfish}, and provides additional useful grid search capabilities.

For very large bandwidths and very many spectral lines, the problem of identifying and cataloging line imperfections essentially becomes a book-keeping and continuum assignment problem.  \texttt{blase} and \texttt{starfish} provide different strategies for orchestrating the line-mismatch identification procedure, with each route having tradeoffs depending on the application.

\section{Conceivable Technical Improvements}\label{technicalInnovations}

Blas\'e already performs very well under a wide range of cloning and transfer-learning tasks.  However, some precision applications may demand even-more-strenuous performance than what the current implementation can accomodate.  Here we describe some of these technical improvements, their design, and/or mention some science case they may unlock.

\subsection{Exact instead of pseudo Voigt Profile}
We currently employ the pseudo-Voigt profile for its low computational cost.  We have a prototype exact-Voigt-Hjerting implementation following \citet{2022ApJS..258...31K}.  We coarsely estimate that moving to this exact-Voigt implementation could decrease some residual regions by $\sim 30\%$, while increasing the computational cost by more than $10\times$ over the existing pseudo-Voigt approximation.  The exact-Voigt-Hjerting implementation still outbids the higher cost of a direct numerical convolution of a Gaussian and Lorentzian profile.

\subsection{Addressing the pseudo-continuum with Gaussian Process regression}

We currently assume the input spectra are adequately normalized to the continuum.  We have a few options to relax this assumption.  We could  simply tune the $\mathsf{P}$ term that represents the wavelength-dependent pre-factor to Equation \ref{equation1}.  Tuning $\mathsf{P}$ would correct for large-scale imperfections in the otherwise-fixed continuum flattening procedure.  This change would be easy and effective, but has some challenge with model selection and flexibility: how to set the polynomial order to avoid over- and under-fitting.  Gaussian Processes (GPs) offer many advantages for continuum fitting \citep{czekala15}.  In short, a GP-likelihood relaxes the assumption that the continuum has been perfectly normalized, in favor of the more realistic statement ``the continuum has been coarsely normalized, with some characteristic-but-as-yet-unknown correlation and scale length and amplitude of the imperfections''.  That statement translates to the following modification to Equation \ref{simpleLikelihood}:

\begin{eqnarray}
    \mathcal{L} =  \frac{1}{2}\mathsf{R^\intercal} \mathsf{C}^{-1} \mathsf{R} +\frac{1}{2}\ln{\det{\mathsf{C}}} \label{GPLikelihood}
\end{eqnarray}

\noindent where we introduce the covariance matrix $\mathsf{C}$, with its associated kernel and collection of typically 2-3 parameters.  We anticipate that this GP likelihood would have the greatest impact on stars with significant band-heads and line-blanketing: spectra with a so-called ``pseudo-continuum''.  M-dwarfs and brown dwarfs fall into this challenging category.

The main demerit of moving to a GP-likelihood is computational cost.  Fortunately, a few efficient autodiff-aware implementations of GPs exist. The \texttt{celerite} algorithm \emph{celerit\'e} \citep{2017AJ....154..220F} has an exact backpropation implementation \citep{2018RNAAS...2...31F} that scales linearly with the number of data points.  The \emph{celerit\'e} algorithm does not currently have a PyTorch implementation.  The \texttt{GPyTorch} framework \citep{2018arXiv180911165G} has a large category of approximate and exact GPs that could be straight-forwardly dropped into \texttt{blase}.  Even still these GPs could increase the computation cost by of order $10\times$.

\subsection{Minibatches and Stochastic Gradient Descent}\label{minibatches}

Currently, each training epoch sees the entire dataset, a setup dubbed \emph{full-batch} gradient descent.
An alternative scheme allows training with only a portion of the entire dataset at a time in \emph{minibatches}.  The massive data volumes in modern Neural Network applications cannot fit into the GPU memory, so minibatches are a necessity.  Our meager 1 MB dataset can easily fit into the GPU memory, but our model can be large if we have a large number of pixels or lines or both.
So while minibatches may not be required due to data size limitations they may be useful for particularly large models.
Minibatches also act as a form of regularization, the principal source of stochasticity in the Stochastic Gradient Descent algorithm, which tends to have better convergence than full-batch Gradient Descent \citep{2016arXiv160904747R}.

We experimented with minibatches by assembling and evaluating only a portion of the dense $\bar{\bm{F}}$ matrix at a time, in minibatches. The choice to evaluate only a portion of lines at a time would mean the model is inaccurately evaluated at all wavelength pixels. Instead, we choose to evaluate all lines, but only on a random subset $N_{\mathrm{batch}}$ of the total pixels $N_s$, so that the model can eventually converge to exact at those points. All lines are allowed to update at each glimpse of a minibatch, but many lines with cores far from minibatch pixels will provide only weak information about how the loss scalar changes for their parameters.

Overall minibatches as implemented above performed worse than the sparse implementation, with both lower accuracy and slower computation time.

\subsection{Broad lines and advanced lineshapes}

Some lines---such as those arising from hydrogen, sodium, potassium, and others---have extremely broad line wings, approaching larger than the $\sim6000$ pixels we allocate for the sparse implementation. These special lines should be handled separately from the weak lines, both from a computational performance perspective and an accuracy perspective.

Extremely broad lines will exhibit truncation effects if the sparse window is small compared to the line wing size. The truncation effects will look like tophat functions severing the asymptotic wings, imbuing artificial step function kinks in the emulated spectrum. We can afford to increase the sparse window on a few, say $N_{broad}\sim20$ of the broadest spectral lines. We then construct and evaluate the entire dense matrix for those lines: $\sim 330\;000 \times 20$. The number of FLOPS in each category scales as about 6 Million for the 20 broad and dense lines versus about 36 Million for the sea of about 7000 narrow and sparse lines, depending on the exact choices for wing cuts and the number of lines.

One could introduce advanced lineshapes for these $\sim20$ broad lines, perturbing the Voigt line-wings with a smooth wavelength-dependent correction term $\mathsf{G}$:

\begin{eqnarray}
    \mathsf{\tilde{V}(\bm{\lambda})} &=& \mathsf{V} \odot \mathsf{G}\\
    \mathsf{G} &=& 1 + (e^{\alpha_j} - 1) \cdot \mathcal{S}\left(\frac{|\bm{\lambda}-\lambda_{c,j}| - \lambda_{t, j}}{b_j}\right)
\end{eqnarray}

\noindent where $\odot$ is again the element-wise product (\emph{a.k.a} Hadamard product), $\mathcal{S}$ is the sigmoid function, and we have introduced three new tunable parameters for each of the $j$ broad lines; $\lambda_t$ is the truncation wavelength, $b$ is a scale parameter for how slowly or how rapidly in wavelength-space the transition from non-Lorentzian proceeds, and $\alpha$ is a possibly negative stretch parameter that controls whether the line wing is sub- or super- Lorentzian.

This functional form has a few advantages. It is smooth. The smoothness of the transition is controlled by a tunable parameter, $b$. It can handle either sub- or super-Lorentzian shapes.  In the limit $\lim_{a\to0} \mathsf{G}$, the lineshape becomes exactly Lorentzian. The sigmoid is efficiently implemented in PyTorch.
Finally, it enforces that the perturbation only produces absorption and not emission profiles.

\subsection{Using native line lists rather than clones}
For many practitioners, the choice to clone precomputed synthetic models in the first place may seem roundabout: ``Why not just use the line lists?''.  Adopting the line lists would have many advantages: it would provide chemical and molecular provenance tags.  Metadata associated with the quality of the atomic and molecular data could be used to assign physics-informed regularization.  Many other benefits would effortlessly accrue from adopting the native line lists.  The \texttt{FAL} project (Cargile et al. \emph{in prep}) follows such a principled prescription.

As already emphasized, there exist at least a few demerits of adopting the line lists, and therefore supporting the \emph{blas\'e} strategy.  First, these line lists need to undergo expensive multi-level radiative transfer calculations in order to obtain their amplitudes, so adopting the line lists would mean a laborious and computationally expensive pursuit simply to get close to what has already been computed.  Second, as the effective temperature scales to ultracool dwarfs (Figure \ref{fig_Nlines_vs_teff}) the number of lines sky-rockets, tending towards the billions for T-dwarfs.  The methane line list alone \citep{2020ApJS..247...55H} represents a prohibitive data volume.  The ExoJAX and Radis \citep{2019JQSRT.222...12P,2021JQSRT.26107476V} libraries offer a breakthrough solution to the voluminous line list problem.  Even still,  \emph{blas\'e} deals with the less pure but more practical ``pseudo line'' that gets closer to the astronomical observables anyways, and offers a middle ground between the extremes of interpretability and performance.

\subsection{Wavelength dependent limb darkening}
Currently the extrinsic model step possesses up to four parameters: the $v\sin{i}$ and $RV$, and 2 optional parameters for limb-darkening.  These four parameters may adequately parameterize a star with a uniform stellar disk.  Extremely high signal-to-noise-ratio spectra of rapidly rotating stars may require additional flexibility.  The limb darkening is generally wavelength-dependent, and so a pan-chromatic spectrum may require a different limb darkening from the blue end to the red end.  The limb darkening may instead depend on physical properties of the spectral line formation, such as physical depth of formation, and so the extent of limb-darkening may jump haphazardly from line-to-line-to-line, rather than as a predictably smooth function across wavelength.  \texttt{Blas\'e} could be built to handle such a seemingly pathological scenario by adding a vector of limb darkening parameters, one for each line.  One would have to regularize the fits with some typical limb darkening and a heuristic penalty for departures from this mean.

\end{document}